\documentclass[aps,prd,onecolumn,nofootinbib,superscriptaddress,amsmath,amssymb,notitlepage]{revtex4-1}
\usepackage{graphicx}
\usepackage{dcolumn}
\usepackage[colorlinks,citecolor=blue,bookmarks]{hyperref}
\usepackage[normalem]{ulem}

\usepackage{multirow}
\usepackage{longtable}

\newcommand{\EOS}{equation of state}
\newcommand{\EOSs}{equations of state}
\newcommand{\EOSx}{equation-of-state}
\newcommand{\NS}{neutron star}
\newcommand{\NSx}{neutron-star}
\renewcommand{\citet}[1]{Ref.~\cite{#1}}

\newcommand{\mnras}{Mon. Not. R. Astron. Soc.}
\newcommand{\apjl}{Astrophys. J. Lett.}
\newcommand{\araa}{Annu. Rev. Astron. Astrophys.}
  
\begin{document}

\title{Inferring neutron star properties from GW170817 with universal relations}

\author{Bharat Kumar}
\email{bharatiop@iucaa.in}
\affiliation{Inter-University Centre for Astronomy and Astrophysics, Post Bag 4, 
Ganeshkhind, Pune, 411007, India}
\author{Philippe Landry}
\email{landryp@uchicago.edu}
\affiliation{Enrico Fermi Institute and Kavli Institute for Cosmological Physics, \linebreak The University of Chicago, 5640 South Ellis Avenue, Chicago, Illinois, 60637, USA}

\date{\today}

\begin{abstract}
Because all neutron stars share a common equation of state, tidal deformability constraints from the compact binary coalescence GW170817 have implications for the properties of neutron stars in other systems.
Using equation-of-state insensitive relations between macroscopic observables like moment of inertia ($I$), tidal deformability ($\Lambda$) and stellar compactness, we derive constraints on these properties as a function of neutron-star mass based on the LIGO-Virgo collaboration's canonical deformability measurement, $\Lambda_{1.4} = 190^{+390}_{-120}$.
Specific estimates of $\Lambda$, $I$, dimensionless spin $\chi$, and stellar radius $R$ for a few systems targeted by radio or X-ray studies are extracted from the general constraints. 
We also infer the canonical neutron-star radius as $R_{1.4} = 10.9^{+1.9}_{-1.5}$ km at 90$\%$ confidence. 
We further demonstrate how a gravitational-wave measurement of $\Lambda_{1.4}$ can be combined with independent measurements of neutron-star radii to tighten constraints on the tidal deformability as a proxy for the equation of state. We find that GW170817 and existing observations of six thermonuclear bursters in low-mass X-ray binaries jointly imply $\Lambda_{1.4} = 196^{+92}_{-63}$ at the 90$\%$ confidence level.
\end{abstract}

\maketitle

\section{Introduction}\label{sec:intro}

The macroscopic properties of neutron stars, like masses, radii, and tidal deformabilities, are highly sensitive to the nuclear microphysics of the stellar interior. Nonetheless, relations between pairs of these observables are often remarkably \emph{insensitive} to internal structure: while a \NS's properties depend individually on the extreme-matter equation of state, certain combinations of them effectively do not.
Several nearly \EOSx~independent relations among \NSx~observables have been studied under the designation of \emph{approximate universal relations} (see \citet{YagiILQreview} for a review).
These include I-Love-Q relations between the moment of inertia $I$, the tidal deformability (or Love number) $\Lambda$ and the rotational quadrupole moment $Q$ \citep{YagiILQscience,YagiILQ}; effective no-hair relations among the lowest few multipole moments \citep{Pappas,Stein,Yagi_nohair}; and binary Love relations linking the tidal deformabilities $\Lambda_{1,2}$ of the members of a binary system \citep{Yagi_BiLoveLett,Yagi_BiLove}.
Neutron-star universality has been proposed as a tool for constraining observationally inaccessible properties \citep{YagiILQreview}, enhancing gravitational-wave parameter estimation \citep{Yagi_BiLove,Chatziioannou}, reducing uncertainty in electromagnetic radius measurements \citep{Psaltis,Baubock}, and testing general relativity (see \citet{Doneva} for a review), among other applications.

The \EOSx~insensitivity of the relations connecting a single \NS's various properties is thought to arise from an emergent symmetry in strongly gravitating stars \citep{Yagi_why}. This kind of universality can be used to translate a measurement of e.g.~a \NS's tidal deformability into a determination of the same star's moment of inertia with percent-level error \citep{YagiILQscience}.
In conjunction with the assumption that all {\NS}s share a common \EOS---a consequence of fundamental nuclear many-body physics---one can moreover establish approximate universal relations between the properties of \emph{different} {\NS}s, like the binary Love relations.
The relations need not be restricted to members of a binary system; measurements of one \NS~have implications for the properties of all other cold, $\beta$-equilibrated {\NS}s in the universe.

Indeed, identical universal relations with comparably small dispersion hold whether the {\NS}s are composed of hadronic, quark \citep{YagiILQscience} or two-phase hybrid hadron-quark \citep{Paschalidis,Wei} matter.
We caution, however, that neutron-star universality is violated by nonbarotropic \EOSs, such as those describing young, hot {\NS}s \citep{Martinon,Marques}, and by the presence of strong stellar magnetic fields, like those associated with magnetars \citep{Haskell}; inferences derived from universal relations are therefore valid for weakly magnetized isolated {\NS}s long after formation and binary {\NS}s long before merger. 
Furthermore, universality appears to be broken by certain non-standard \EOSs~with strong phase transitions \citep{Bandyopadhyay,Han,Lau,Annala17}.
Disagreement between universal-relation based predictions and direct measurements of astrophysical {\NS}s could thus be a signature of such \EOSs.

The universal I-Love relation and a specially adapted binary Love relation were combined by \citet{Landry_pulsar} to infer the moment of inertia of PSR J0737-3039A, the primary component of the double pulsar, with $\approx 30\%$ accuracy based on tidal deformability constraints from GW170817 \citep{LVC_GW170817,LVC_eos}.
We extend this technique to make general inferences about the properties of {\NS}s, placing bounds on tidal deformability, moment of inertia and radius $R$ as a function of stellar mass $M$ via universal binary Love, I-Love and I-compactness relations.\footnote{A different universal relation has been used elsewhere in conjunction with GW170817 to constrain the maximum mass of nonrotating {\NS}s \citep{Rezzolla}.}
We take the constraint $\Lambda_{1.4} = 190^{+390}_{-120}$ (median and symmetric 90$\%$ confidence interval) on the canonical deformability of a $1.4\,M_{\odot}$ \NS~established by \citet{LVC_eos} as our primary observational input.
Their study assumed a common \EOS~\citep{Chatziioannou} and reprised the initial GW170817 parameter estimation of \citet{LVC_GW170817}, which found $\Lambda_{1.4} \leq 800$ at 90$\%$ confidence, assuming small \NSx~spins, by performing a Bayesian analysis of the gravitational-wave strain data recorded by Advanced LIGO \citep{LIGO} and Virgo \citep{Virgo}.
The original parameter estimation results were also combined with priors on the \EOS~from parametric piecewise-polytrope \citep{Annala18} and perturbative QCD \citep{Most2018} models to obtain the constraints $\Lambda_{1.4} \in [120,1504]$ (allowing for first-order phase transitions) and $\Lambda_{1.4} > 375$ (95$\%$ confidence, assuming purely hadronic composition), respectively.
Similarly, \citet{LandryEssick} used updated parameter estimation results from \citet{LVC_properties} with a broad non-parametric \EOSx~prior to find $\Lambda_{1.4} = 160^{+448}_{-113}$ (maximum \textit{a posteriori} and highest-posterior-density 90$\%$ confidence interval).\footnote{In the remainder of the paper, quoted error bars refer to symmetric 90$\%$ confidence intervals about the median unless otherwise specified.}
We present general tidal deformability, moment of inertia, and radius bounds associated with these constraints for comparison with those derived from \citet{LVC_eos}.

Refs.~\cite{De,LVC_properties} also measured \NSx~tidal deformability with GW170817, but they reported the chirp deformability $\tilde{\Lambda}$ rather than the canonical deformability $\Lambda_{1.4}$. 
The former is a mass-weighted average of the tidal deformabilities of the {\NS}s involved in the coalescence, and is therefore specific to the event GW170817; the latter is a generic constraint on the \EOS~that is more easily incorporated in our universal relations.
Likewise, multimessenger parameter estimation studies of GW170817 and its electromagnetic counterpart, combining gravitational-wave and kilonova observations, yielded intriguing constraints on $\tilde{\Lambda}$ \citep{RadiceDai,Radice,Coughlin2018}, in addition to other macroscopic observables \citep{Bauswein}.
The conclusions of Refs.~\cite{De,LVC_properties} are similar to those of \citet{LVC_eos}, favoring a relatively soft \EOS, while the multimessenger analyses indicate a preference for a somewhat larger tidal deformability, corresponding to a slightly stiffer \EOS~consistent with the findings of \citet{Annala18}.

Besides gravitational-wave measurement of the tidal deformability, masses and radii have been measured for a variety of pulsars and binary {\NS}s via radio and X-ray astronomy.
However, only a handful of \emph{simultaneous} mass and radius measurements exist \citep{OzelFreire}.
Even the most precise of these, obtained by fitting spectra for accretion-powered thermonuclear bursts on the surface of {\NS}s in low-mass X-ray binaries, may be affected by substantial systematic errors \citep{MillerLamb}.
Nonetheless, we extract radius estimates for six bursters studied by \citet{Ozel} from our general constraints, and find that they are consistent with the corresponding electromagnetic measurements. 
In the cases we consider, the universal-relation based constraints on $R$ turn out to be more precise than the direct radius measurements themselves, after accounting for the uncertainty in the burster masses.

Additionally, we estimate moments of inertia for a few short-period double {\NS}s whose relativistic periastron advance may be measurable with next-generation radio observatories, like the Square Kilometre Array \citep{SKA}.
Future direct measurements of $I$ can be compared to these gravitational-wave predictions to test the universality of the \EOS~\citep{Landry_pulsar}.
We perform a similar moment-of-inertia calculation for millisecond pulsars of known mass. 
Using their measured angular frequencies $\Omega$, we compute their dimensionless spins $\chi := cI\Omega/GM^2$ to be $O(0.1)$.
For the fastest spinning pulsars in double \NS~systems, we find instead $\chi \sim 0.01$, in keeping with conventional expectations \citep{Damour,Hannam}.

In anticipation of more accurate \NSx~radius measurements from the NICER observatory \citep{NICER}, we demonstrate how the binary Love, I-Love and I-compactness relations can be combined into an effective $R(M,\Lambda_{1.4})$ relation that is insensitive to the \EOS.  
This derived relation can be employed to place multimessenger constraints on tidal deformability using gravitational waves from binary \NS~mergers in conjunction with radius measurements from X-ray binaries.
Taking simultaneous mass and radius measurements for the six thermonuclear bursters as our input, we tighten the GW170817-derived bounds on canonical deformability to $\Lambda_{1.4} = 196^{+92}_{-63}$, assuming all the observations are equally reliable. 
The constraint's improved precision, relative to previous results, reinforces existing observational support for a particularly soft \EOS.
Obtaining this type of multimessenger constraint from universal relations is simpler than performing a joint Bayesian analysis and does not require modeling the \EOS.\footnote{Note added: A Bayesian analysis of this kind---but focused on the stellar radius, rather than the tidal deformability---is presented in \citet{Fasano}, which appeared shortly after completion of this paper.}

We describe our universal-relation based inference of \NS~properties below. The \EOSs~used to compute the relations are presented in Sec.~\ref{sec:uni}, with the piecewise polytrope representation we adopt detailed in Appendix~\ref{sec:pwp}. The binary Love, I-Love and I-compactness fits are introduced in Secs.~\ref{sec:bilove}-\ref{sec:ic}. Sec.~\ref{sec:scheme} explains our inference method. The results of the inference for general \NSx~observables, as well as for specific systems, are presented in Secs.~\ref{sec:gen} and \ref{sec:inf}, respectively. Multimessenger constraints on \NS~tidal deformability are calculated in Sec.~\ref{sec:joint}. Lastly, we discuss our findings in Sec.~\ref{sec:disc}.

\section{Universal relations} \label{sec:uni}

To infer the tidal deformabilities, moments of inertia, and radii of astrophysical neutron stars from gravitational-wave observations, we require universal relations linking each of these properties to the canonical deformability deduced through Bayesian parameter estimation \citep{LVC_GW170817,LVC_eos}.
The desired I-Love, binary Love, and I-compactness relations have been computed elsewhere, but for consistency in modeling the error in the fits we recompute the latter two here with the fiducial set of \EOSs~used in \citet{Landry_pulsar}.
We also specialize the binary Love relation of \citet{Yagi_BiLove,Yagi_BiLoveLett} to our purposes.
We therefore briefly recapitulate our choice of \EOSs, and our calculation of sequences of neutron-star observables, before presenting the specific fits employed for the universal relations.

\begin{figure}
\centering
\includegraphics[width=0.66\columnwidth]{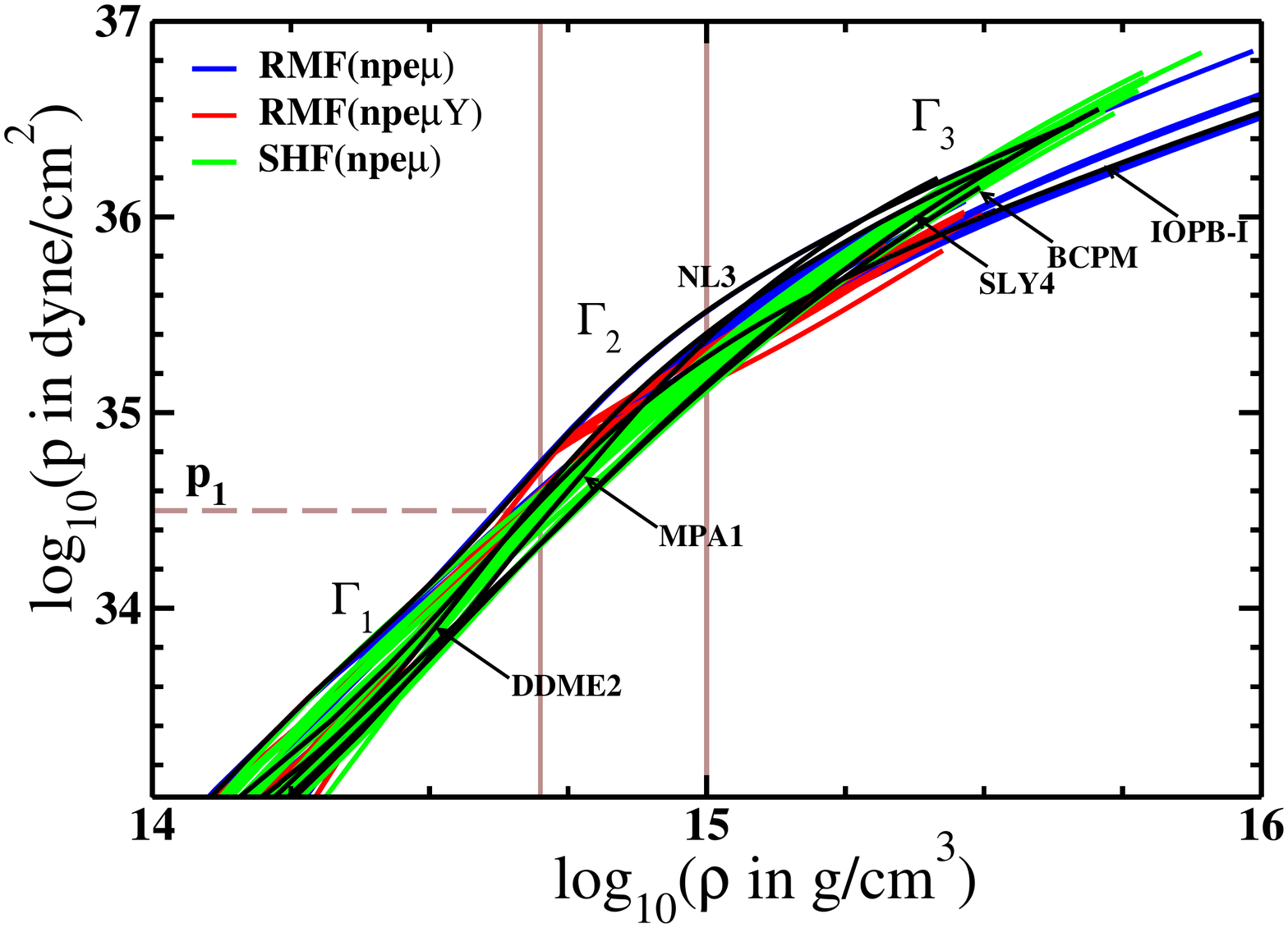}
\caption{Pressure $p$ as function of rest-mass energy density $\rho$ for the RMF and SHF \EOSs. The \EOSs~are colored by type and composition, with a few labeled explicitly for reference. The dividing densities of the three-segment piecewise polytrope representation we adopt for calculations with the \EOSs~(see Appendix~\ref{sec:pwp}) are indicated with vertical lines, and the piecewise polytrope parameters are shown schematically.}
\label{fig:eos}
\end{figure}

\citet{Landry_pulsar} computed the I-Love relation and a binary Love relation between $\Lambda_{1.4}$ and PSR J0737-3039A's tidal deformability using a collection of 53 unified \EOSs~from relativistic mean-field (RMF) theory and Skyrme-Hartree-Fock (SHF) theory.
The \EOSs, plotted in Fig.~\ref{fig:eos}, are consistent with studies of the bulk properties of finite nuclei and infinite nuclear matter near nuclear saturation density, as well as the observational lower bound on the \NSx~maximum mass \citep{Antoniadis}, which we conservatively take as $1.93\,M_{\odot}$.
The set includes RMF models with hyperonic $npe\mu Y$ matter, in addition to hadronic RMF and SHF $npe\mu$-matter models, and spans a wide range in stiffness and phenomenological behavior. 
Although none of these \EOSs~include quark matter, as per the Introduction, we expect the universal relations for purely hadronic stars to hold to the same level of accuracy for quark stars and two-phase hadron-quark hybrids.
A complete listing of the \EOSs~is given in Sec.~2 of \citet{Landry_pulsar}, and we adopt the same set for our calculations here.

The \NSx~observables are determined by integrating the equations of stellar structure for a choice of \EOS~and central density.
Specifically, the Tolman-Oppenheimer-Volkoff equations \citep{Tolman,Oppenheimer} fix the stellar mass and radius, Hartle's slow-rotation equation \citep{Hartle} sets the moment of inertia, and the field equation for the quadrupolar tidal perturbation governs the tidal deformability \citep{Hinderer08}.
A stable sequence of {\NS}s is obtained from successive choices of central density such that the resulting masses span from 1 to $1.93\,M_{\odot}$.
(For consistency, we truncate every sequence at $1.93 \, M_{\odot}$, even if the \EOS~can support a more massive star.)
For the purpose of these integrations, we adopt a piecewise polytrope representation of the \EOS~\citep{Read}.
This phenomenological parameterization is commonly used in astrophysics and gravitational-wave astronomy because it accurately reproduces with four parameters the \NS~properties one would calculate from a tabulated version of the \EOS.
We determine the accuracy of the piecewise polytrope fits to the RMF and SHF \EOSs~in Appendix~\ref{sec:pwp}, and list the best-fit parameter values in Table~\ref{tb:params1}.

\subsection{Binary Love relation} \label{sec:bilove}

\begin{figure}
\centering
\includegraphics[width=0.66\columnwidth]{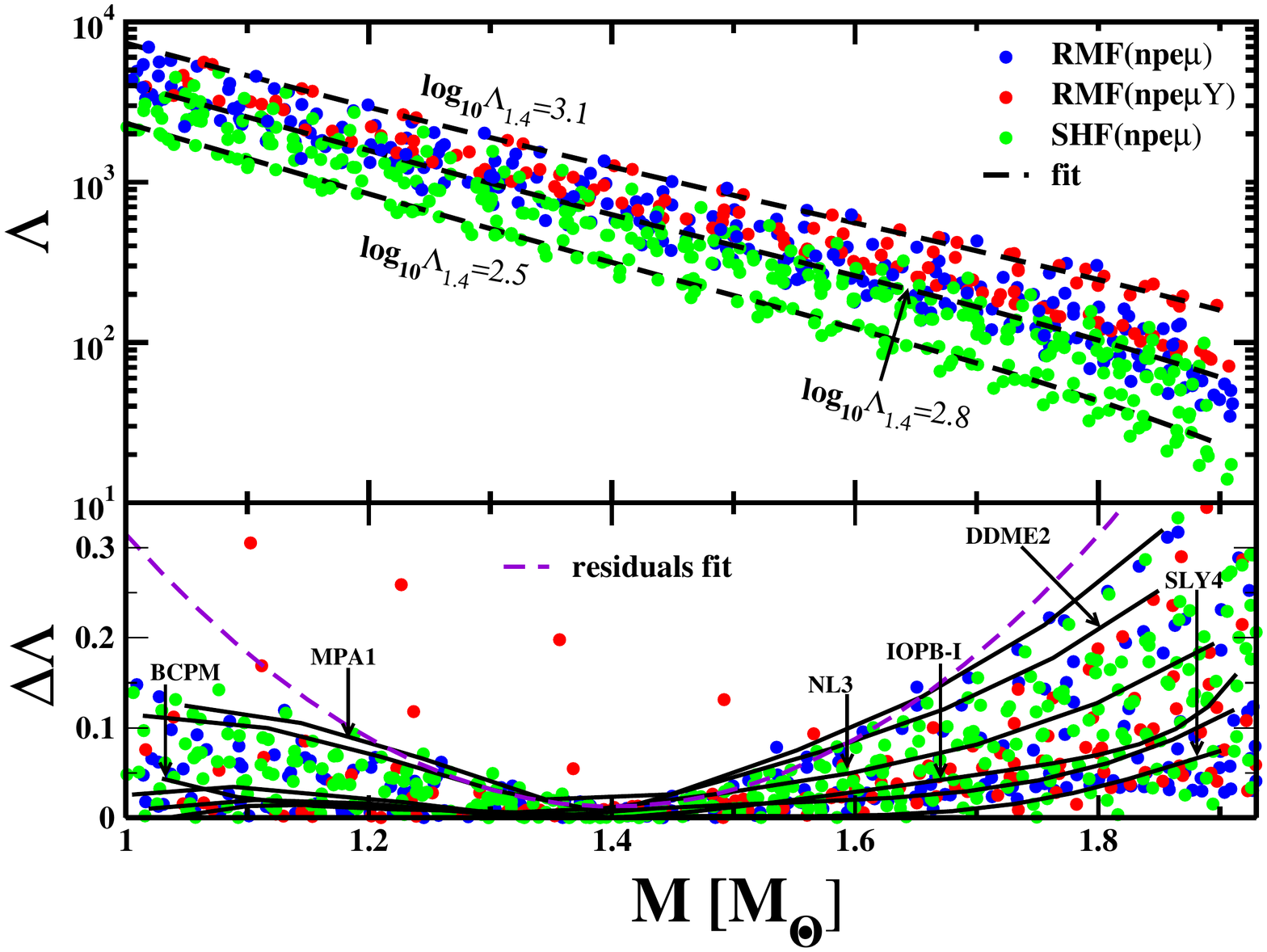}
\caption{Binary Love relation calculated with our set of 53 \EOSs. The black dashed lines in the upper panel are selected $\Lambda_{1.4}=\text{constant}$ slices of the three-dimensional fit \eqref{bilove} to the sequences of $(M,\Lambda_{1.4},\Lambda)$ data. The fit and data are projected into the $M$-$\Lambda$ plane for display purposes only. The fit residuals are plotted in the lower panel, where the purple dashed curve approximates the maximum residuals in accordance with Eq.~\eqref{biloveres}.}
\label{fig:bilove}
\end{figure}

We calculate a binary Love relation between the tidal deformability of a $1.4\,M_{\odot}$ star and that of a star of mass $M$ by performing a three-dimensional fit to $(M, \Lambda_{1.4}, \Lambda)$ data for a stable sequence of neutron stars with each of the 53 equations of state described above.
Expanding in canonical deformability and stellar mass, we posit a functional form

\begin{equation} \label{bilove}
\log_{10} \Lambda = \sum_{m=0}^{4} \sum_{n=0}^{1} a_{mn} M^m (\log_{10} \Lambda_{1.4})^n
\end{equation}
for the relation and perform a least-squares fit for $a_{mn}$.
The resulting coefficients are listed in Table~\ref{tb:coeff}.
Projections of the fit surface into the $M$-$\Lambda$ plane are superimposed on the underlying \NSx~data in  Fig.~\ref{fig:bilove}, which also shows the fit residuals

\begin{equation} \label{biloveres}
\Delta \Lambda = \frac{|\Lambda-\Lambda_{\text{fit}}|}{\Lambda_{\text{fit}}} .
\end{equation}
The residuals are calculated in the full three-dimensional space, but are projected into the $M$-$\Lambda$ plane in the plot.
The maximum residuals as a function of mass can be approximated by

\begin{equation} \label{resfit}
\Delta \Lambda (M) = b_0 + b_1 M + b_2 M^2 ,
\end{equation}
with the coefficients $b_n$ given in Table~\ref{tb:coeff}; for simplicity, we suppress the $\Lambda_{1.4}$-dependence of the residuals in our representation of the dispersion. 
The function $\Delta \Lambda(M)$ is used to model the errors in the fit \eqref{bilove}.
Specifically, denoting the best-fit tidal deformability relation from Eq.~\eqref{bilove} as $\Lambda_{\text{fit}}$, we take $\Delta \Lambda \, \Lambda_{\text{fit}}$ to be half the width of the symmetric, two-sided 90$\%$ confidence interval of a Gaussian distribution

\begin{equation} \label{biloveprob}
P(\Lambda|M,\Lambda_{1.4}) = \frac{1}{\sqrt{2\pi{\sigma_{\Lambda}}^2}} \exp\left[-(\Lambda-\Lambda_{\text{fit}})^2/2{\sigma_{\Lambda}}^2 \right]
\end{equation}
centered on $\Lambda_{\text{fit}}$ that characterizes the uncertainty in the relation due to its \emph{approximately} universal nature. 
Here, $\sigma_{\Lambda}(M) = \Delta \Lambda(M) \, \Lambda_{\text{fit}}/1.645$ is the standard deviation derived from the fractional errors $\Delta \Lambda(M)$.
The fractional errors are $O(1\%)$ near $1.4\,M_{\odot}$ and rise to $\approx 50\%$ at the high-mass edge of the relation. 

\setlength{\tabcolsep}{6pt}
\begin{table}
\centering
\caption{Coefficients of the fits \eqref{bilove}, \eqref{resfit}, \eqref{ilove} and \eqref{ic} for the binary Love, I-Love and I-compactness relations.} \label{tb:coeff}
\begin{tabular}{lcccc}
\hline \hline
 \multicolumn{2}{c}{$\Lambda(M,\Lambda_{1.4})$} & $\Delta \Lambda(M)$ & $\bar{I}(\Lambda)$ & $C(\bar{I})$  \\ \hline
 $a_{00} = -9.4469 \phantom{\times 10^{-1}}$ & $a_{01} = \phantom{-}4.6152 \phantom{\times 10^{-1}} $ & $b_0 = \phantom{-}3.7152$ & $c_0 = \phantom{-}6.5022 \times 10^{-1}$ & $d_0 = \phantom{-}4.8780 \times 10^{-2}$ \\
 $a_{10} = \phantom{-}3.9702 \times 10^{1}$ & $a_{11} = -1.2226 \times 10^{1}$ & $b_1 = -5.2874$ & $c_1 = \phantom{-}5.8594 \times 10^{-2}$ & $d_1 = -4.2829 \times 10^{-1}$ \\
$a_{20} = -4.9173 \times 10^{1}$ & $a_{21} = \phantom{-}1.4214 \times 10^{1}$ & $b_2 = \phantom{-}1.8876$ & $c_2 = \phantom{-}5.1749 \times 10^{-2}$ & $d_2 = \phantom{-}1.2468 \phantom{\times 10^{-1}} $ \\
$a_{30} = \phantom{-}2.4937 \times 10^{1}$ & $a_{31} = -7.1134 \phantom{\times 10^{-1}}$ & - & $c_3 = -3.6321 \times 10^{-3}$ & $d_3 = -9.0716 \times 10^{-1}$ \\
$a_{40} = -4.7288 \phantom{\times 10^{-1}}$ & $a_{41} = \phantom{-}1.3416 \phantom{\times 10^{-1}}$ & - & $c_4 = \phantom{-}8.5909 \times 10^{-5}$ & $d_4 = \phantom{-}2.3302 \times 10^{-1}$ \\
\hline \hline
\end{tabular}
\end{table}

The binary Love relation \eqref{bilove} is similar to the one introduced by Refs.~\cite{Yagi_BiLove,Yagi_BiLoveLett}, but is specially adapted to our purpose.
While the original binary Love relation effectively links $\Lambda_1(M_1)$ and $\Lambda_2(M_2)$ via the mass ratio $M_2/M_1$, assuming a common equation of state, ours essentially sets $\Lambda_1 = \Lambda_{1.4}$ by fixing $M_1 = 1.4\,M_{\odot}$, and accordingly we use $M_2$ itself in place of the mass ratio $M_2/(1.4 \, M_{\odot})$.
Moreover, Refs.~\cite{Yagi_BiLove,Yagi_BiLoveLett} use the combinations $\Lambda_s = (\Lambda_1+\Lambda_2)/2$ and $\Lambda_a = (\Lambda_1-\Lambda_2)/2$ in place of the individual tidal deformabilities to improve the universality of the fit.
Doing the same would unnecessarily complicate our inference, as a closed-form expression for $\Lambda_2$ cannot be obtained from a log-log polynomial fit for $(\Lambda_s,\Lambda_a)$.
In any case, the dispersion in our modified binary Love relation is nearly as small as in the original formulation.

\subsection{I-Love relation} \label{sec:ilove}

We adopt the I-Love relation from Eq.~(7) of \citet{Landry_pulsar} directly, as it was computed with the same set of equations of state considered here. The coefficients of the log-log polynomial fit

\begin{equation}\label{ilove}
\log_{10}{\bar{I}} = \sum_{n=0}^{4} c_n \left( \log_{10}{\Lambda} \right)^n
\end{equation}
for the dimensionless moment of inertia $\bar{I} := c^4 I/G^2 M^3$ as a function of tidal deformability are given in Table~\ref{tb:coeff}.
The fit, the ($\Lambda$,$\bar{I}$) data, and the residuals

\begin{equation} \label{iloveres}
  \Delta \bar{I} = \frac{|\bar{I}-\bar{I}_{\text{fit}}|}{\bar{I}_{\text{fit}}}
\end{equation}
can be seen in Fig.~4 of \citet{Landry_pulsar}.
The maximum residuals are approximately constant over the relevant range of $\Lambda$, amounting to no more than 0.6$\%$ error.
We therefore take this value to define the half-width of the 90$\%$ confidence interval of the Gaussian uncertainty in the fit, modeled in the same manner as above, such that

\begin{equation} \label{iloveprob}
P(\bar{I}|\Lambda) = \frac{1}{\sqrt{2\pi{\sigma_{\bar{I}}}^2}} \exp\left[-(\bar{I}-\bar{I}_{\text{fit}})^2/2{\sigma_{\bar{I}}}^2 \right] ,
\end{equation}
with $\sigma_{\bar{I}} = \Delta \bar{I} \, \bar{I}_{\text{fit}}/1.645$.

\subsection{I-compactness relation} \label{sec:ic}

Our universal relation between the dimensionless moment of inertia and the stellar compactness $C := GM/c^2 R$ is based on a similar one from \citet{Breu}.
Quasi-universal I-compactness relations predate the I-Love-Q relations in the literature \citep{Ravenhall,Bejger,LattimerMoI,Baubock}, but generally exhibit a lesser degree of \EOSx~independence \citep{Chan,YagiILQreview}.
\citet{Breu} discovered that the relation's universality could be enhanced by using the normalization $\bar{I} = c^4 I/G^2 M^3$ for the dimensionless moment of inertia, as in Refs.~\cite{YagiILQscience,YagiILQ}, rather than the conventional definition $I/M R^2$.
Hence, theirs is the version of the I-compactness relation we calculate here; however, we fit for the inverse relation, namely $C(\bar{I})$.

\begin{figure}
\centering
\includegraphics[width=0.66\columnwidth,]{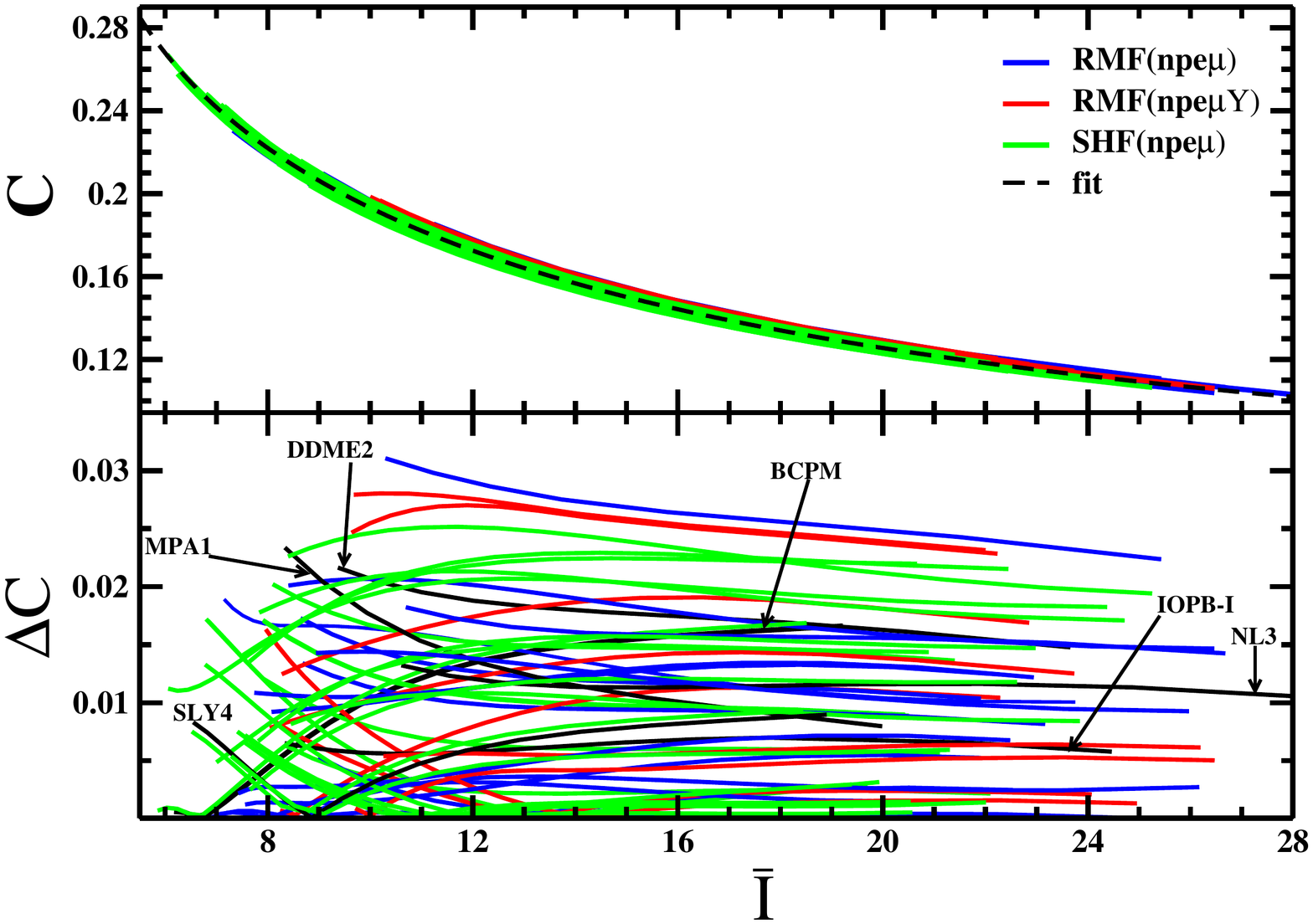}
\caption{I-compactness relation calculated with our set of 53 \EOSs. The fit \eqref{ic} is shown as a black dashed line, and the fit residuals are displayed in the lower panel.}
\label{fig:ci}
\end{figure}

Taking ($\bar{I}$, $C$) data for our stable sequences of {\NS}s, we perform a least-squares fit to the model

\begin{equation}\label{ic}
C = \sum_{n=0}^{4} d_n \left( \log_{10}{\bar{I}} \right)^{-n} ,
\end{equation}
displaying the resulting coefficients in Table~\ref{tb:coeff}.
The fit and the residuals

\begin{equation} \label{icres}
  \Delta C = \frac{|C-C_{\text{fit}}|}{C_{\text{fit}}}
\end{equation}
are shown alongside the underlying \NS~data in Fig.~\ref{fig:ci}.
The maximum residuals are roughly constant as a function of $\bar{I}$, so we take the maximum value of $3\%$ to define the half-width of the 90$\%$ confidence interval about the mean of the Gaussian distribution describing the error in the relation,

\begin{equation} \label{icprob}
P(C|\bar{I}) = \frac{1}{\sqrt{2\pi{\sigma_{C}}^2}} \exp\left[-(C-C_{\text{fit}})^2/2{\sigma_{C}}^2 \right]
\end{equation}
with $ \sigma_C = \Delta C \, C/1.645$.

\section{Inference scheme} \label{sec:scheme}

Equipped with the distributions \eqref{biloveprob}, \eqref{iloveprob} and \eqref{icprob} describing the probabilistic mappings defined by the universal relations, we can translate a gravitational-wave measurement of $\Lambda_{1.4}$ into constraints on {\NS}s' tidal deformabilities, moments of inertia, spins and radii. Our inference of these observables is described in general terms here; we apply it to the observational input from GW170817 in the following section.

We suppose that $P(\Lambda_{1.4} |\,\text{GW})$, the posterior probability distribution for the canonical deformability given a gravitational-wave observation $\text{GW}$, is known. 
The target of our inference is taken to be a \NS~for which a mass posterior $P(M|\,\text{EM})$ is available from electromagnetic observations $\text{EM}$. 
(The case of general constraints on \NSx~properties, absent a specific target system, is treated separately below.) 
The posterior distributions from the independent gravitational-wave and electromagnetic measurements serve as priors for our inference of the target's properties.

Firstly, the posterior distribution $P_{\Lambda}(\Lambda|\,\text{EM},\text{GW})$ for the target's tidal deformability, conditioned on the gravitational-wave and electromagnetic observations, is computed by marginalizing the binary Love relation over the two priors:

\begin{equation} \label{plambda}
P_{\Lambda}(\Lambda|\,\text{EM},\text{GW}) = \int P(\Lambda |\,M,\Lambda_{1.4}) P(M |\,\text{EM}) P(\Lambda_{1.4} |\,\text{GW}) \, dM \, d\Lambda_{1.4} .
\end{equation}
The posterior distribution for the target's dimensionless moment of inertia can then be calculated via the I-Love relation as

\begin{equation} \label{pibar}
 P_{\bar{I}}(\bar{I}|\,\text{EM},\text{GW}) = \int P(\bar{I}|\,\Lambda) P_{\Lambda}(\Lambda|\,\text{EM},\,\text{GW}) \, d\Lambda
 \end{equation} 
by marginalizing over the tidal deformability. 
The posteriors for the moment of inertia $I = G^2 \bar{I} M^3/c^4$ and the dimensionless spin $\chi = G \bar{I} M \Omega/c^3$ follow by a change of variables and a marginalization over mass:

\begin{align} \label{pi}
P_I(I |\,\text{EM},\text{GW}) &= \frac{c^4}{G^2}\int \frac{P_{\bar{I}}(c^4 I/G^2 M^3 |\,\text{EM},\text{GW}) P(M |\,\text{EM})}{M^{3}} \, dM , \\
P_{\chi}(\chi |\,\text{EM},\text{GW}) &= \frac{c^3}{G}\int \frac{P_{\bar{I}}(c^3 \chi/G M \Omega |\,\text{EM},\text{GW}) P(M |\,\text{EM})}{M \Omega} \, dM .  \label{pchi}
\end{align} 
When inferring $\chi$, we assume that the {\NS}'s rotational frequency $\Omega$ is known exactly.

From Eq.~\eqref{pibar}, we can also infer the posterior distribution for the target's compactness $C = GM/c^2 R$ through

\begin{equation} \label{pc}
P_C(C |\,\text{EM},\text{GW}) = \int P(C|\,\bar{I}) P_{\bar{I}}(\bar{I}|\,\text{EM},\text{GW}) \, d\bar{I} ,
\end{equation}
which makes use of the I-compactness relation and leads immediately to an inference of the target's radius via

\begin{equation} \label{pr}
P_R(R |\,\text{EM},\text{GW}) = \frac{G}{c^2} \int \frac{P_C(GM/c^2R |\,\text{EM},\text{GW}) P(M |\,\text{EM})}{R^2} M \, dM .
\end{equation}
In the event that the target's mass is known exactly, $P(M |\,\text{EM})$ reduces to a Dirac delta function in $M$ and the mass marginalizations are trivial.

Given the posterior distributions \eqref{plambda}, \eqref{pi}, \eqref{pchi} and \eqref{pr}, we can compute the median value of $\Lambda$, $I$, $\chi$ and $R$ for the target star and extract symmetric confidence intervals for each observable.
General constraints on \NSx~properties, rather than inferences for a specific target system, can also be calculated by dispensing with the mass marginalizations altogether and computing the posterior distributions as a function of mass based on the gravitational-wave observation alone, i.e.~

\begin{align} \label{plambdam}
P_{\Lambda_M}(\Lambda|\,M;\text{GW}) = \int P(\Lambda |\,M,\Lambda_{1.4}) P(\Lambda_{1.4} |\,\text{GW}) \, d\Lambda_{1.4} , \\
P_{\bar{I}_M}(\bar{I}|\,M;\text{GW}) = \int P(\bar{I}|\,\Lambda) P_{\Lambda_M}(\Lambda|\,M;\text{GW}) \, d\Lambda , \label{pim} \\
P_{R_M}(R |\,M;\text{GW}) = P_{C_M}(GM/c^2R |\,M;\text{GW}) \, GM/c^2R^2 . \label{prm}
\end{align} 
Here, we have defined $P_{C_M}(C |\,M;\text{GW}) := \int P(C|\,\bar{I}) P_{\bar{I}_M}(\bar{I}|\,M;\text{GW})\, d\bar{I}$. 
By calculating confidence intervals about the median \textit{a posteriori} for each value of mass in the domain, we can place constraints on the $M$-$\Lambda$, $M$-$\bar{I}$, and $M$-$R$ relations that govern all old, cold {\NS}s in the universe.

\section{Implications of GW170817 for neutron star properties} \label{sec:results}

We apply the inference described above to astrophysical {\NS}s, using GW170817---and, specifically, the measurement $\Lambda_{1.4} = 190^{+390}_{-120}$ from \citet{LVC_eos}---as our observational gravitational-wave input. 
However, since only the median and symmetric $90\%$ confidence interval for $\Lambda_{1.4}$ were reported in \citet{LVC_eos}, we must model the full posterior distribution $P(\Lambda_{1.4}|\,\text{GW})$. In order to preserve the asymmetry evident in the confidence interval, we choose to represent it as a generalized beta prime distribution 

\begin{equation} \label{model}
P(\Lambda_{1.4}|\,\text{GW}) = \frac{p \,\Gamma(\alpha+\beta)}{q\,\Gamma(\alpha)\Gamma(\beta)} \left(\frac{\Lambda_{1.4}}{q}\right)^{\alpha p-1}\left[1-\left(\frac{\Lambda_{1.4}}{q}\right)^p\right]^{-\alpha-\beta}
\end{equation}
with parameters $p=1$, $q=0.934$, $\alpha = 2.856$ and $\beta = 191.509$, where $\Gamma(z)$ is the gamma function. 
With these parameter selections, the distribution has the same symmetric $90\%$ confidence interval as implied by the gravitational-wave measurement, and its median of 198 is only shifted mildly relative to the actual value. 
Thus, our model for $P(\Lambda_{1.4}|\,\text{GW})$ closely reproduces the features of the measurement reported in \citet{LVC_eos}.

Using this posterior probability distribution, we first infer general constraints on the $M$-$\Lambda$, $M$-$\bar{I}$, and $M$-$R$ relations, and then extract specific bounds for individual {\NS}s of interest. 
Because our universal relations' fits and errors are based on data for $M \in [1, 1.93] \, M_{\odot}$, we focus on {\NS}s with (median) $M \leq 1.93 \, M_{\odot}$ in this paper to avoid extrapolation insofar as possible.

\subsection{General constraints} \label{sec:gen}

\begin{figure}
\centering
\includegraphics[width=0.66\columnwidth]{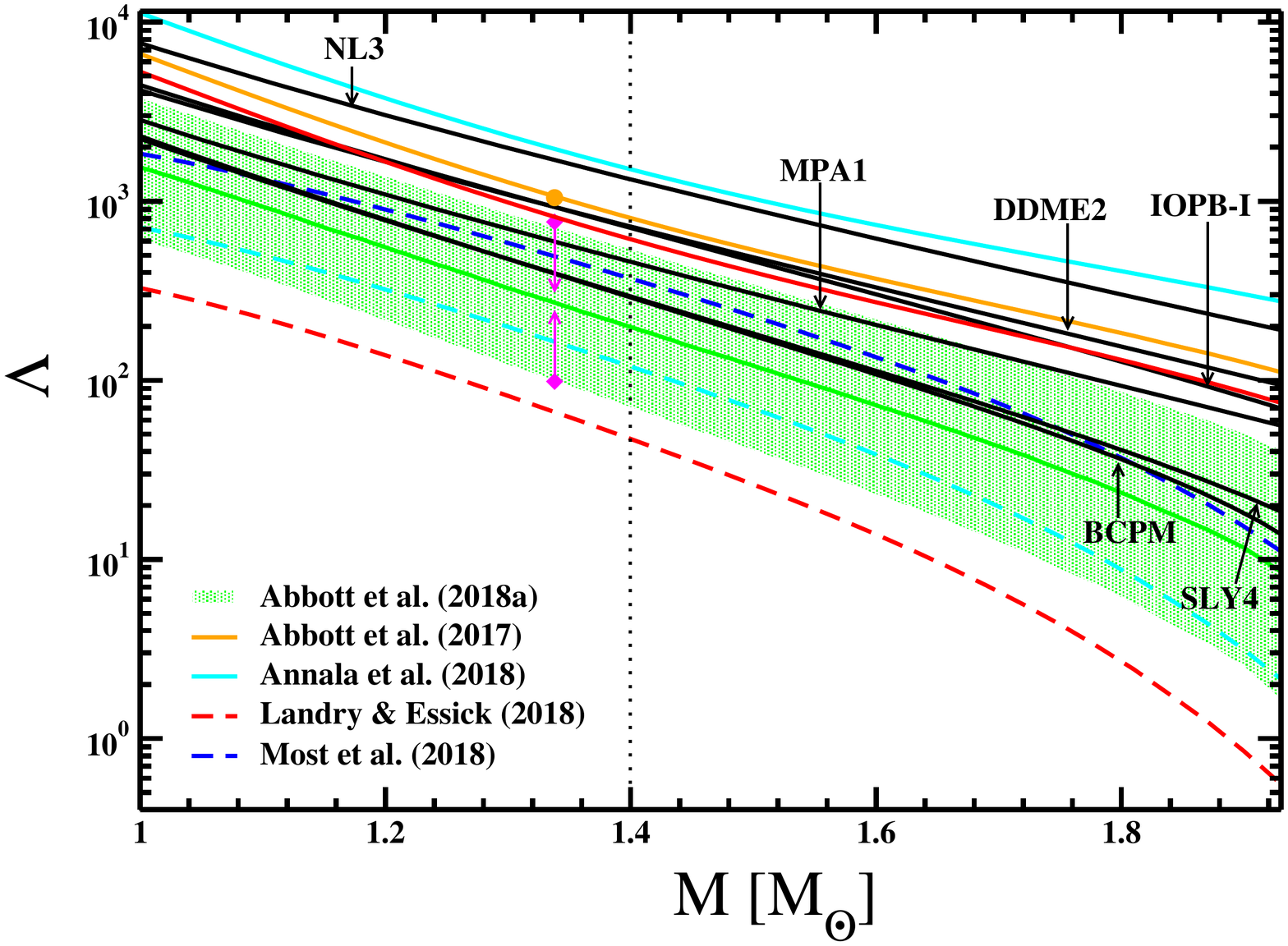}
\caption{Constraints on the mass--tidal-deformability relation $\Lambda(M)$ from GW170817 and the universal relations. The green line and shaded region show the median and symmetric $90\%$ confidence interval derived from \citet{LVC_eos}'s $\Lambda_{1.4}$ measurement. Upper (respectively lower) bounds stemming from alternative constraints are indicated by solid (dashed) colored lines; the input $\Lambda_{1.4}$ constraints can be read off from the $\Lambda(M)$ curves at $M = 1.4\,M_{\odot}$ (dotted vertical line). $\Lambda(M)$ relations for a few reference \EOSs~are shown in black. The tidal deformability inferred for the $1.338\,M_{\odot}$ PSR J0737-3039A by \citet{Landry_pulsar} on the basis of \citet{LVC_eos}'s (\citet{LVC_GW170817}'s) canonical deformability measurement is indicated with the pink error bars (orange point).}
\label{fig:mlambda}
\end{figure}

\begin{figure}
\centering
\includegraphics[width=0.66\columnwidth]{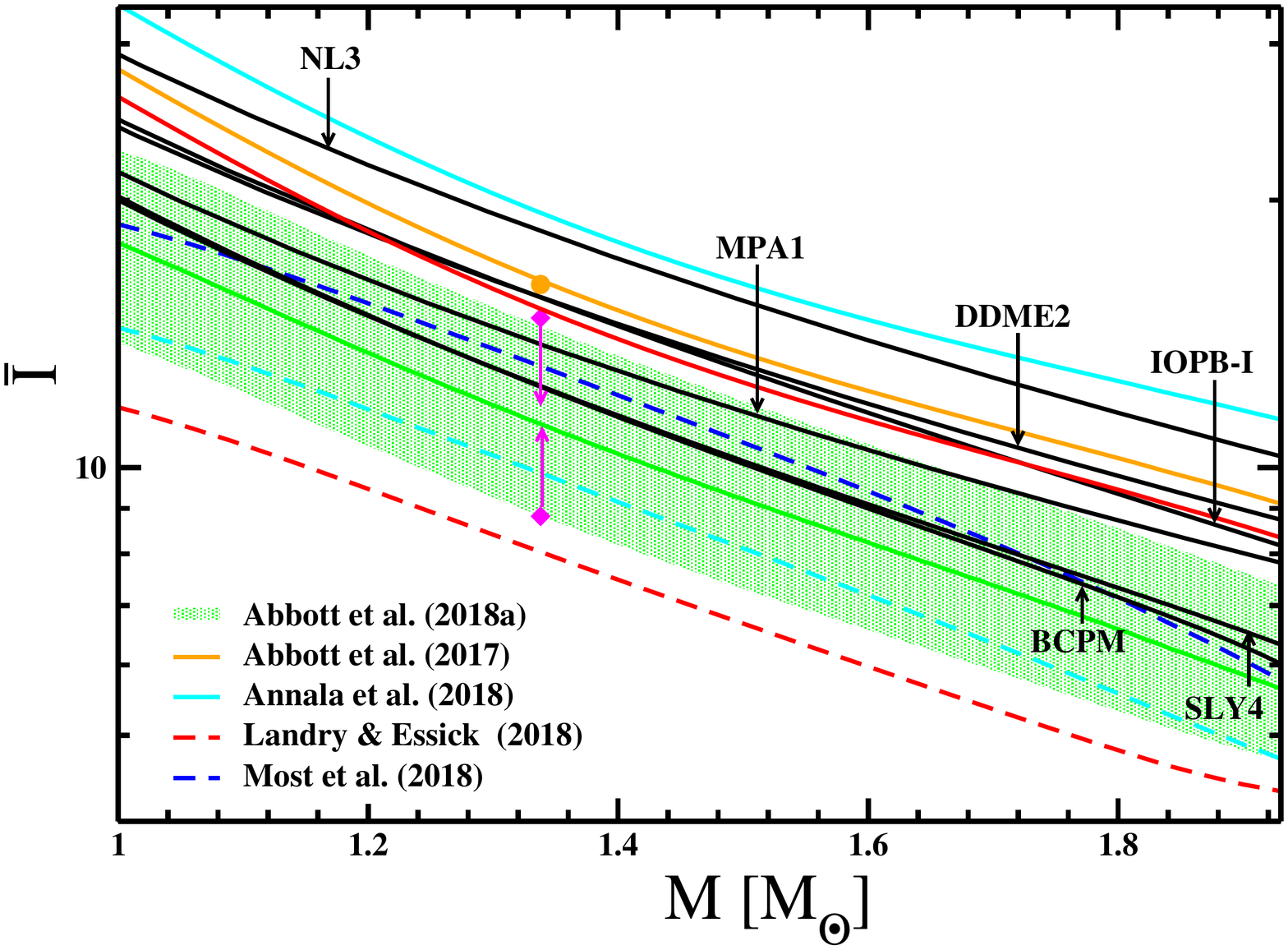}
\caption{Constraints on the mass--moment-of-inertia relation $\bar{I}(M)$ from GW170817 and the universal relations. $\bar{I}(M)$ relations for a few reference \EOSs, as well as the dimensionless moment of inertia inferred for the double pulsar by \citet{Landry_pulsar}, are also shown.}
\label{fig:mi}
\end{figure}

\begin{figure}
\centering
\includegraphics[width=0.66\columnwidth]{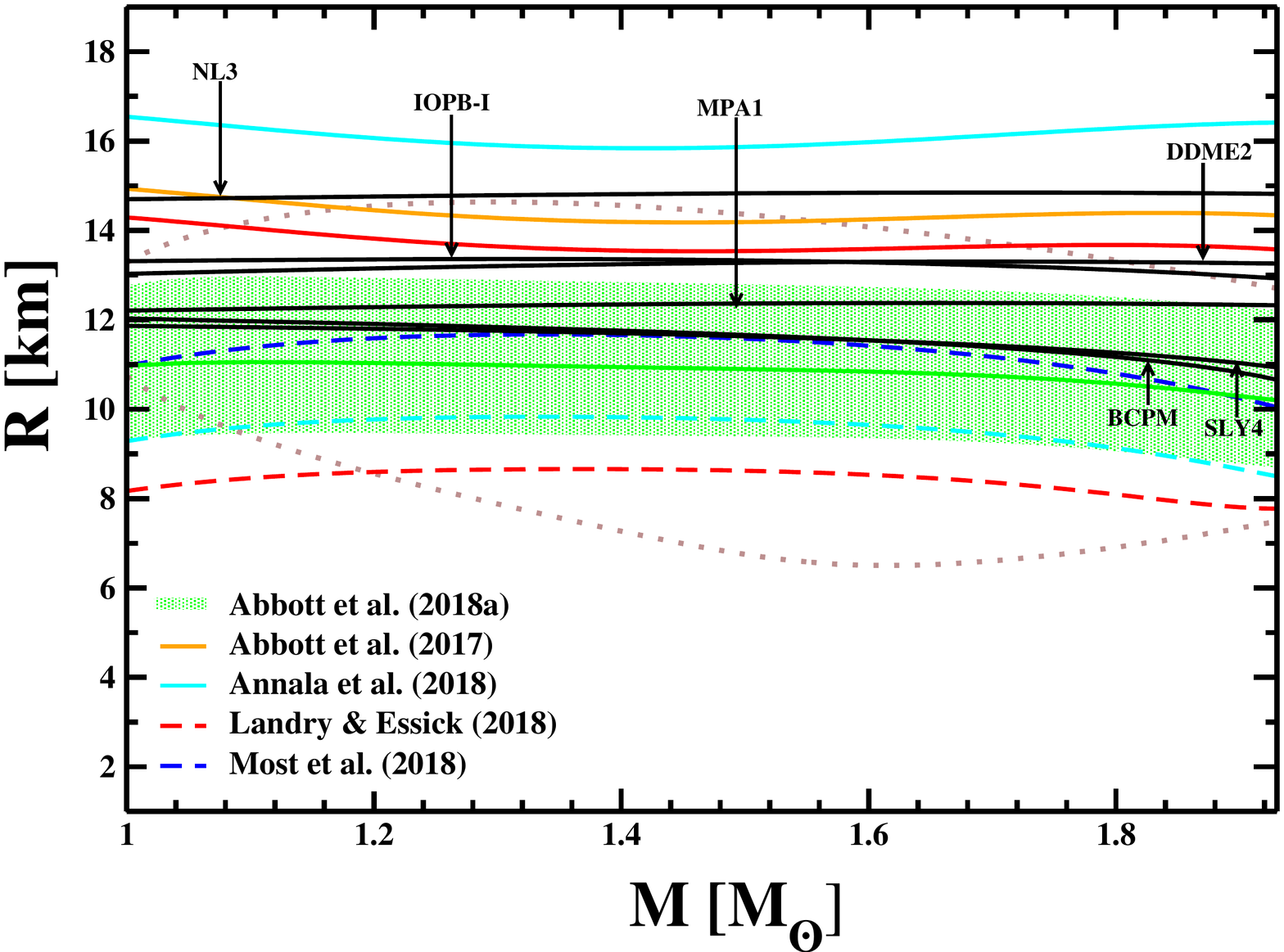}
\caption{Constraints on the mass-radius relation $R(M)$ from GW170817 and the universal relations. $R(M)$ relations for a few reference \EOSs~are also shown. The posterior 90$\%$-credible contour from the simultaneous mass and radius measurement of EXO 1745-248 is overplotted in brown, demonstrating the consistency of the universal-relation based inference with electromagnetic observations of {\NS}s.}
\label{fig:mr}
\end{figure}

Following Eqs.~\eqref{plambdam}-\eqref{prm}, we calculate, as a function of mass, symmetric $90\%$ confidence intervals about the median for each of the neutron-star properties attainable from the GW170817 $\Lambda_{1.4}$ measurement by way of the universal relations.
The resulting constraints on $\Lambda(M)$, $\bar{I}(M)$ and $R(M)$ are plotted in Figs.~\ref{fig:mlambda}-\ref{fig:mr}.
The canonical deformability measurement maps to the colored band with decreasing slope in the $M$-$\Lambda$ plane seen in Fig.~\ref{fig:mlambda}, reflecting the fact that $\Lambda$ is a monotonically decreasing function of mass.
We observe that several of the stiffer reference \EOSs, e.g.~NL3, DDME2, and IOPB-I, lie outside the 90$\%$ confidence region, in keeping with the preference found by other studies \citep{LVC_GW170817,De,LVC_eos,LVC_properties} for a relatively soft \EOS. 
Similarly, the $\Lambda(M)$ constraint transforms to the colored band in the $M$-$\bar{I}$ plane shown in Fig.~\ref{fig:mi}. 
Its decreasing slope reflects the monotonicity of $\bar{I}(M)$ for realistic \EOSs, and the same stiff models are disfavored.

The corresponding $R(M)$ constraint is depicted in Fig.~\ref{fig:mr}. 
The median $M$-$R$ relation reveals that \NSx~universality imposes near-constancy of the radius over the mass range of interest. 
The colored region of the plot excludes radii larger than 13.0 km and less than 8.7 km at 90$\%$ confidence for stars with $M \in [1,1.93]\,M_{\odot}$.
Evaluating the constraint at $M = 1.4 \, M_{\odot}$, GW170817 implies $R_{1.4} = 10.9^{+1.9}_{-1.5}$ km for the canonical radius.
This value is compatible with upper bounds of $\approx 13$-14 km computed via \EOSx~modeling \citep{Annala18,Nandi,Fattoyev_GW170817,Most2018} or a universal chirp-deformability--radius relation \citep{Raithel2018}.
It also overlaps with the result $R_{1.4} = 12.2^{+1.0}_{-0.8}$ km obtained by \citet{RadiceDai}'s joint gravitational-wave and electromagnetic parameter estimation for GW170817.

To illustrate how the inferred bounds on the properties as a function of mass depend on the choice of priors and assumptions made in the initial parameter estimation for $\Lambda_{1.4}$, in Figs.~\ref{fig:mlambda}-\ref{fig:mr} we also show the general constraints stemming from \citet{LVC_GW170817} ($\Lambda_{1.4} \leq 800$, without the common \EOS~assumption), \citet{Annala18} ($120 \leq \Lambda_{1.4} \leq 1504$, modeling the \EOS~as a piecewise polytrope), \citet{Most2018} ($\Lambda_{1.4} > 375$, modeling the \EOS~via perturbative QCD calculations), and \citet{LandryEssick} ($\Lambda_{1.4} = 160^{+448}_{-113}$, modeling the \EOS~with a Gaussian process).
Since these results are only shown for comparative purposes, we do not perform the full inference described in Sec.~\ref{sec:scheme}. 
Rather, we simply map each $\Lambda_{1.4}$ constraint through the best-fit universal relations, accounting for uncertainty by inflating upper and lower bounds by a factor of the fractional error in the fit. 
In this way, we obtain conservative estimates of the alternative constraints' implications for \NSx~properties. 
As can be seen, \citet{Annala18}'s upper bound is stiff enough to allow all the reference \EOSs. Meanwhile, depending on the analysis, the constraint's lower bound excludes a varying fraction of the region compatible with \NSx~universality.
We note that the maximum \textit{a posteriori} from \citet{LandryEssick} is omitted in the plots, as it is  similar to the median from \citet{LVC_eos}.

\subsection{Individual neutron stars} \label{sec:inf}

Next, we extract constraints on the properties of specific {\NS}s of interest---primarily pulsars with well-measured masses from electromagnetic observations---following Eqs.~\eqref{plambda}-\eqref{pr}.
In the literature, universal relations have been proposed as a tool for improving the precision of a measurement of \NSx~radius, spin, or moment of inertia \citep{Maselli,YagiILQreview,Bhat}; however, until recently \citep{Landry_pulsar}, the application has always involved translation of one observable (e.g.~$\Lambda$) to another (e.g.~$I$) for the same system.
Here we use GW170817 to infer the properties of individual {\NS}s in other systems.
We compute their tidal deformabilities, moments of inertia and radii using the $\Lambda_{1.4}$ constraint from \citet{LVC_eos}.

\subsubsection{Double neutron stars} \label{sec:dns}

We begin by inferring the tidal deformability, moment of inertia, and radius for the pulsar component of several short-period double \NSx~systems. 
Double neutrons stars in tight binaries are good candidates for an electromagnetic measurement of the stellar moment of inertia if radio pulses from one member of the system are detectable, as they can be used to determine the post-Keplerian parameters of the orbit with great precision \citep{KramerWex}.
In particular, a sufficiently precise measurement of the system's relativistic periastron advance can distinguish the part due to spin-orbit coupling, which is proportional to the spin of the pulsar; knowledge of its angular frequency can then be used to extract the moment of inertia, which depends sensitively on the \EOS.
No electromagnetic \NSx~$I$ measurements exist at present, but they may be feasible with the Square Kilometre Array and other next-generation radio observatories \citep{Kehl}.

The best-studied candidate for a future moment of inertia measurement is the double pulsar, PSR J0737-3039 \citep{Burgay03,Lyne04}.
Its 1.338$\, M_{\odot}$ primary component's moment of inertia was estimated by \citet{Landry_pulsar} as $1.15^{+0.38}_{-0.24} \times 10^{45}$ g cm$^2$ based on GW170817 and universal relations. 
We revisit the calculation here, ignoring the uncertainty of less than one part in $10^{-3}$ in the pulsar's mass---i.e.~taking $P(M|\,\text{EM})=\delta(M-1.338\,M_{\odot})$---when using Eq.~\eqref{pi}. 
Despite using a slightly different binary Love relation than \citet{Landry_pulsar}, we find a nearly identical moment of inertia constraint, $I = 1.16^{+0.33}_{-0.25}\times 10^{45}$ g cm$^2$. 
Furthermore, we infer the pulsar's tidal deformability to be $269^{+439}_{-170}$, and its radius as $11.0^{+1.9}_{-1.5}$ km. 
This radius value is no different, within uncertainty, than that of a canonical 1.4$\, M_{\odot}$ star.

PSR J1946+2052 is another especially promising candidate for a moment of inertia measurement, since it resides in the tightest double \NSx~system discovered to date \citep{Stovall}; it is also the fastest-spinning pulsar with a \NSx~companion that will merge within a Hubble time. 
However, given its relatively recent discovery, the pulsar's mass has not yet been determined with precision---only an upper bound of 1.31$\, M_{\odot}$ exists. 
Nonetheless, if we model $P(M|\,\text{EM})$ as flat for $M \in [1.0,1.3] \, M_{\odot}$, we are able to estimate its moment of inertia as $0.96^{+0.37}_{-0.26}\times 10^{45}$ g cm$^2$ by marginalizing over the mass uncertainty.

In Table~\ref{tb:dns}, we report the inferred properties of these and several other pulsars in double \NSx~systems, such as PSR B1913+16, the Hulse-Taylor pulsar \citep{HulseTaylor}.
As with PSR J0737-3039A, we take their masses to be known exactly for the purpose of the calculation, except for the aforementioned case of PSR J1946+2052. The errors reported in the table therefore account for the approximate nature of the universal relations and the uncertainty in the $\Lambda_{1.4}$ measurement from GW170817.
Although the pulsar masses are clustered in the small range of $\approx 1.3$-$1.6 \, M_{\odot}$, the inferred 90$\%$ confidence intervals for the tidal deformabilities are distributed over an order of magnitude.
The $\Lambda$ uncertainties are typically lopsided, with larger error bars on the upper side, because the monotonically decreasing function $\Lambda(M)$ behaves roughly like $1/M$, tending to a constant value at large $M$.
For the moments of inertia, we find that they are typically constrained by GW170817 to $\approx 30\%$ accuracy, with median values of $\sim 1 \times 10^{45}$ g cm$^2$. 
Given the weak mass dependence of the radius for $M \in [1,1.93] \, M_{\odot}$, the median radius for nearly all the pulsars in Table~\ref{tb:dns} is $11.0$ km.

\begin{table}
\centering
\caption{Inferred properties of pulsars in double neutron star systems. The tidal deformability, moment of inertia, radius, and dimensionless spin are calculated via universal relations from the $\Lambda_{1.4}$ constraint of \citet{LVC_eos}. Orbital periods, masses and rotational frequencies are drawn from the listed references. The measurement uncertainty  of no more than $\pm 1$ in the last digit of $M$ is ignored for the purpose of the inference, with the exception of PSR J1946+2052, for which we assume a flat probability distribution for $M \in [1,1.3] \,M_{\odot}$.}
\label{tb:dns}
\begin{tabular}{lclcccccc} \hline \hline
Pulsar & $P_{\text{orb}}$ [d] & $M$ $[M_{\odot}]$ & $\Omega$ [rad s$^{-1}$] & Reference & $\Lambda$ & $I$ $[10^{45}~\text{g cm}^2]$ & $R$ [km] & $\chi$  \\ \hline
B1534+12 &  0.421 &  \phantom{$<\;$}1.333 & 165.76 & \cite{Fonseca} & $276^{+449}_{-174}$ & $1.15^{+0.33}_{-0.24}$ & $11.0^{+1.9}_{-1.5}$ & $0.012^{+0.004}_{-0.003}$  \\
B1913+16 & 0.323 & \phantom{$<\;$}1.438 & 106.44 & \cite{Weisberg} & $163^{+286}_{-106}$ & $1.27^{+0.37}_{-0.27}$ & $10.9^{+1.9}_{-1.5}$ & $0.007^{+0.002}_{-0.002}$ \\
B2127+11C & 0.335 &  \phantom{$<\;$}1.36\phantom{0} & 205.81 & \cite{Jacoby} & $241^{+399}_{-153}$ & $1.19^{+0.34}_{-0.25}$ & $11.0^{+1.9}_{-1.5}$ & $0.015^{+0.004}_{-0.003}$  \\
J0453+1559 & 4.072 &  \phantom{$<\;$}1.56\phantom{0} & 137.24 & \cite{Martinez} & $\phantom{0}89^{+172}_{-60}$ & $1.41^{+0.41}_{-0.29}$ & $10.9^{+1.9}_{-1.5}$ & $0.009^{+0.003}_{-0.002}$  \\
J0737-3039A & 0.102 & \phantom{$<\;$}1.338 & 276.80 & \cite{Kramer} & $269^{+439}_{-170}$ & $1.16^{+0.33}_{-0.25}$ & $11.0^{+1.9}_{-1.5}$ & $0.020^{+0.006}_{-0.004}$ \\
J1756-2251 &  0.320 &  \phantom{$<\;$}1.34\phantom{0} & 220.76 & \cite{Ferdman} & $267^{+435}_{-168}$ & $1.16^{+0.34}_{-0.25}$ & $11.0^{+1.9}_{-1.5}$ & $0.016^{+0.005}_{-0.003}$  \\
J1906+0746 &  0.166 &  \phantom{$<\;$}1.29\phantom{0} & \phantom{0}43.61 & \cite{Leeuwen} & $344^{+542}_{-215}$ &  $1.11^{+0.32}_{-0.24}$ & $11.0^{+1.9}_{-1.5}$ & $0.003^{+0.001}_{-0.001}$  \\
J1946+2052 & 0.078 &  $<\;$1.31\phantom{0} & 370.47 & \cite{Stovall} & $710^{+1516}_{-490}$ & $0.96^{+0.37}_{-0.26}$ & $11.0^{+1.9}_{-1.6}$ & $0.031^{+0.009}_{-0.007}$ \\ 
\hline \hline
\end{tabular}
\end{table}

\subsubsection{Millisecond pulsars} \label{sec:psr}

Precise mass and angular frequency measurements exist for a number of millisecond pulsars thanks to detailed studies of their regular radio pulses.
Here we calculate their moments of inertia as a way to infer their dimensionless spins. 
We focus on a subset of the millisecond pulsars considered in \citet{OzelFreire}, and list their masses, angular frequencies and inferred properties in Table~\ref{tb:psr}.
The subset includes PSR J0437-4715, the closest and brightest pulsar detected to date \citep{Reardon}, and PSR J1713+0747, one of the most precisely timed pulsars \citep{Zhu}.

We model the uncertainty in the pulsars' masses as Gaussian, converting the standard deviations reported in the original references listed in the table to $90\%$ confidence intervals.
With this model for $P(M|\,\text{EM})$, we follow the prescription of Sec.~\ref{sec:scheme} for computing confidence intervals about the median moment of inertia.
Overall, we find that the errors bars on $I$ are slightly larger than for the double {\NS}s in Table~\ref{tb:dns} on account of the broader mass uncertainties for the millisecond pulsars.

Incorporating the pulsars' known angular frequencies, we then infer the stars' dimensionless spins. 
We find that the universal relations permit $\chi$ to be inferred from GW170817 with $\approx 30\%$ accuracy in an approximately \EOSx~independent way. 
The fastest-spinning pulsar we consider, PSR J1909-3744, is found to have $\chi = 0.147^{+0.043}_{-0.031}$.

The astrophysical spin distribution for millisecond pulsars is known to extend up to at least $\chi \sim 0.4$ \citep{Hessels}, while binary {\NS}s that merge within a Hubble time are expected to have much smaller spins $\chi \lesssim 0.05$ \citep{Damour,Hannam,Landry_pulsar}.
Hence, for comparison, we also infer the dimensionless spin for the pulsar components of the double \NS~systems listed in Table~\ref{tb:dns}.
We find that the pulsars of this kind have dimensionless spin $\chi \lesssim 0.04$ at 90$\%$ confidence, while the millisecond pulsars in Table~\ref{tb:psr} have dimensionless spins below $\chi \approx 0.20$.
One could systematize this dimensionless spin inference for all known pulsars to establish a virtually \EOSx~independent upper bound on the spin distribution, whose precision would improve as more gravitational-wave events are detected.

Because the universal relations used here were developed in the context of slowly rotating stellar models, one might suppose that they do not apply to rapidly rotating millisecond pulsars. 
However, Refs.~\cite{Pappas,Chakrabarti} showed that they also hold for stars in rapid uniform rotation,\footnote{Because we evaluate the stability of our \NSx~sequences in the absence of rotation, we are excluding supramassive (i.e.~rotation-stabilized) {\NS}s, for which the universal relations deteriorate at high compactness \citep{Lenka}.} despite earlier claims to the contrary \citep{Doneva_rot}. 
In any case, for stars with moderate rotation ($\chi \sim 0.1$), spin corrections to the moment of inertia are negligible, as they enter at $O(\chi^2) \sim 10^{-2}$.

In addition, we note that our spin analysis depends implicitly on the assumption that the progenitors of GW170817 rotated slowly, with $\chi \leq 0.05$, through the priors adopted in \citet{LVC_eos}'s parameter estimation. 
The low-spin assumption is consistent with dimensionless spin estimates for the fastest-spinning pulsars in double \NSx~systems \citep{Damour,Hannam,Landry_pulsar}. 
However, for a spin inference that is independent of this assumption, one could repeat the calculation with the upper bound $\Lambda \leq 1400$ from \citet{LVC_GW170817}, which instead requires only $\chi \leq 0.89$ \textit{a priori}. 
Indeed, this was done for the double pulsar in Sec.~5 of \citet{Landry_pulsar}.

\begin{table}
\centering
\caption{Inferred properties of millisecond pulsars. The tidal deformability, moment of inertia, radius and dimensionless spin are calculated via universal relations from the $\Lambda_{1.4}$ constraint of \citet{LVC_eos}. Masses and rotational frequencies are drawn from the listed references. The Gaussian errors in $M$ have been converted to the $90\%$ confidence level.}
\label{tb:psr}
\begin{tabular}{lccccccc} \hline \hline
Pulsar & $M$ $[M_{\odot}]$ & $\Omega$ [rad s$^{-1}$] & Reference & $\Lambda$ & $I$ $[10^{45}~\text{g cm}^2]$ & $R$ [km] & $\chi$ \\ \hline
J0437-4715 & $1.44 \pm 0.12$ &  1091.31 & \cite{Reardon} & $163^{+344}_{-116}$ & $1.28^{+0.40}_{-0.29}$ & $10.9^{+1.9}_{-1.5}$ & $0.076^{+0.022}_{-0.016}$ \\
J0751+1807 & $1.64 \pm 0.25$ & 1795.20 & \cite{Desvignes} & $\phantom{0}59^{+227}_{-51}$ & $1.50^{+0.51}_{-0.39}$ &  $10.7^{+1.9}_{-1.6}$ & $0.114^{+0.036}_{-0.026}$ \\
J1713+0747 & $1.31 \pm 0.18$ & 1374.84 & \cite{Zhu} & $310^{+710}_{-232}$ & $1.13^{+0.40}_{-0.30}$ & $11.0^{+1.8}_{-1.5}$ & $0.103^{+0.029}_{-0.023}$ \\
J1802-2124 & $1.24 \pm 0.18$ & \phantom{0}496.79 & \cite{Ferdman10} & $439^{+939}_{-326}$ & $1.05^{+0.38}_{-0.28}$ & $11.0^{+1.8}_{-1.5}$ & $0.038^{+0.011}_{-0.009}$ \\
J1807-2500B & $1.3655 \pm 0.0034$ & 1500.93 & \cite{Lynch12} & $234^{+391}_{-149}$ & $1.19^{+0.35}_{-0.25}$ & $11.0^{+1.9}_{-1.5}$ & $0.109^{+0.032}_{-0.023}$ \\
J1909-3744 & $1.47 \pm 0.05$ &  2131.98 & \cite{Reardon} & $139^{+261}_{-94}$ & $1.31^{+0.38}_{-0.28}$ & $10.9^{+1.9}_{-1.5}$ & $0.147^{+0.043}_{-0.031}$ \\
J2222-0137 & $1.20 \pm 0.23$ & 191.46 & \cite{Kaplan14} & $509^{+1062}_{-397}$ & $1.02^{+0.40}_{-0.29}$ & $10.9^{+1.7}_{-1.6}$ & $0.015^{+0.004}_{-0.003}$ \\ 
\hline \hline
\end{tabular}
\end{table}

\subsubsection{Low-mass X-ray binaries} \label{sec:xray}

Neutron stars in X-ray binaries are the best candidates for electromagnetic radius measurements. Radius estimates for a few systems already exist, although their accuracy is a matter of some debate \citep{MillerLamb}.
The most precise measurements involve thermonuclear bursters in low-mass X-ray binaries; by fitting for the spectrum of the thermal emission, which is related to the burst luminosity by a factor of the surface area, one can determine the radius from the observed flux \citep{OzelFreire}. 
Observations from the NICER mission are expected to place even tighter and more accurate constraints on \NS~radii via pulse profile modeling \citep{OzelNICER}.

For the time being, we focus on six bursters in low-mass X-ray binaries for which simultaneous mass and radius measurements exist \citep{Ozel}. 
In Table~\ref{tb:xray}, we list the median and symmetric $90\%$ confidence intervals for the \NSx~masses and radii extracted from the $M$-$R$ posteriors associated with the electromagnetic observations.\footnote{The mass-radius posteriors are available in tabulated form at \url{http://xtreme.as.arizona.edu/NeutronStars/}.}
(Note that the masses and radii reported in Refs.~\cite{OzelFreire,Ozel} are given instead as maxima \textit{a posteriori} with symmetric uncertainties at the $68\%$ confidence level.)
The confidence intervals are calculated from the marginal distributions $P(M|\,\text{EM}) = \int P(M,R|\,\text{EM}) \,dR$ and $P(R|\,\text{EM}) = \int P(M,R|\,\text{EM}) \,dM$, respectively, with $P(M,R|\,\text{EM})$ constructed from the available posterior samples.
Taking the calculated $P(M|\,\text{EM})$ as our mass prior in Eq.~\eqref{pr}, we obtain a GW170817-based radius estimate for the {\NS}s through the universal relations. 
The inferred radii are consistent with the $R_{\text{EM}}$ values obtained from the direct measurements via $P(R|\,\text{EM})$. 
This can also be seen in Fig.~\ref{fig:mr}, where---as an example---we overlay the 90$\%$ confidence contour of $P(M,R|\,\text{EM})$ for EXO 1745-248 on our $R(M)$ constraints.
In Table~\ref{tb:xray}, besides the inferred radius, we also show the tidal deformability and moment of inertia calculated for each burster. 
We note that, for the thermonuclear bursters considered here, the universal relations and GW170817 actually provide a more precise radius determination at the 90$\%$ confidence level than the direct observations, after marginalizing over the mass posterior $P(M|\,\text{EM})$.

\begin{table}
\centering
\caption{Inferred properties of {\NS}s in low-mass X-ray binaries for which simultaneous mass and radius measurements exist. The tidal deformability, moment of inertia and radius are calculated via universal relations from the $\Lambda_{1.4}$ constraint of \citet{LVC_eos}. Masses and direct radius measurements $R_{\text{EM}}$ are obtained from the $M$-$R$ posteriors associated with \cite{Ozel}, as described in the text.}
\label{tb:xray}
\begin{tabular}{lcccccc} \hline \hline
Neutron star & $M$ $[M_{\odot}]$ & $R_{\text{EM}}$ [km] & $\Lambda$ & $I$ $[10^{45}~\text{g cm}^2]$ & $R$ [km] \\ \hline 
	4U 1608-52 & $1.59^{+0.54}_{-0.47}$ & $10.2^{+3.7}_{-2.7}$ & $74^{+532}_{-72}$ & $1.45^{+0.61}_{-0.53}$ & $10.7^{+1.9}_{-1.7}$ \\
4U 1724-207 & $1.81^{+0.36}_{-0.48}$ & $11.5^{+2.5}_{-2.5}$ & $24^{+291}_{-23}$ & $1.64^{+0.54}_{-0.54}$ & $10.4^{+2.0}_{-1.6}$ \\
4U 1820-30 & $1.76^{+0.44}_{-0.43}$ & $11.2^{+3.2}_{-2.6}$ & $32^{+297}_{-31}$ & $1.60^{+0.56}_{-0.52}$ & $10.5^{+2.0}_{-1.6}$ \\
EXO 1745-248 & $1.60^{+0.36}_{-0.42}$ & $10.3^{+2.7}_{-2.4}$ & $72^{+477}_{-67}$ & $1.45^{+0.56}_{-0.50}$ & $10.7^{+1.9}_{-1.6}$ \\
KS 1731-260 & $1.59^{+0.61}_{-0.62}$ & $10.4^{+3.8}_{-3.4}$ & $67^{+587}_{-65}$ & $1.47^{+0.63}_{-0.57}$ & $10.6^{+1.9}_{-1.7}$ \\
SAX J1748.9-2021 & $1.73^{+0.43}_{-0.56}$ & $11.3^{+2.9}_{-2.9}$ & $37^{+450}_{-36}$ & $1.57^{+0.57}_{-0.59}$ & $10.5^{+1.9}_{-1.7}$ \\
\hline \hline
\end{tabular}
\end{table}

\section{Multimessenger constraints on tidal deformability} \label{sec:joint}

Typical multimessenger probes of the \NSx~\EOS~involve gravitational-wave and electromagnetic measurements of the same system.
However, the universal relations provide a means to translate observations of low-mass X-ray binaries into quantities, like tidal deformabilities, that are normally measured via gravitational waves from binary \NS~mergers.
The independent gravitational-wave and electromagnetic measurements can then be combined to tighten the constraints on the tidal deformability as a proxy for the \EOS.

We use the simultaneous mass and radius measurements for the aforementioned bursters in conjunction with GW170817 to improve knowledge of the canonical deformability, starting with EXO 1745-248 as an example. 
The symmetric 90$\%$ confidence intervals for its mass and radius, calculated from the $M$-$R$ posterior samples associated with the electromagnetic observations, are given in Table~\ref{tb:xray}.
The uncertainty of $\approx 25\%$ in its radius at 90$\%$ confidence is characteristic of the best current measurements; radius measurements with a better level of precision ($\approx 15\%$ at 90$\%$ confidence) are expected from pulse profile modeling with NICER \citep{OzelNICER}.

\begin{figure}
\centering
\includegraphics[width=0.55\columnwidth]{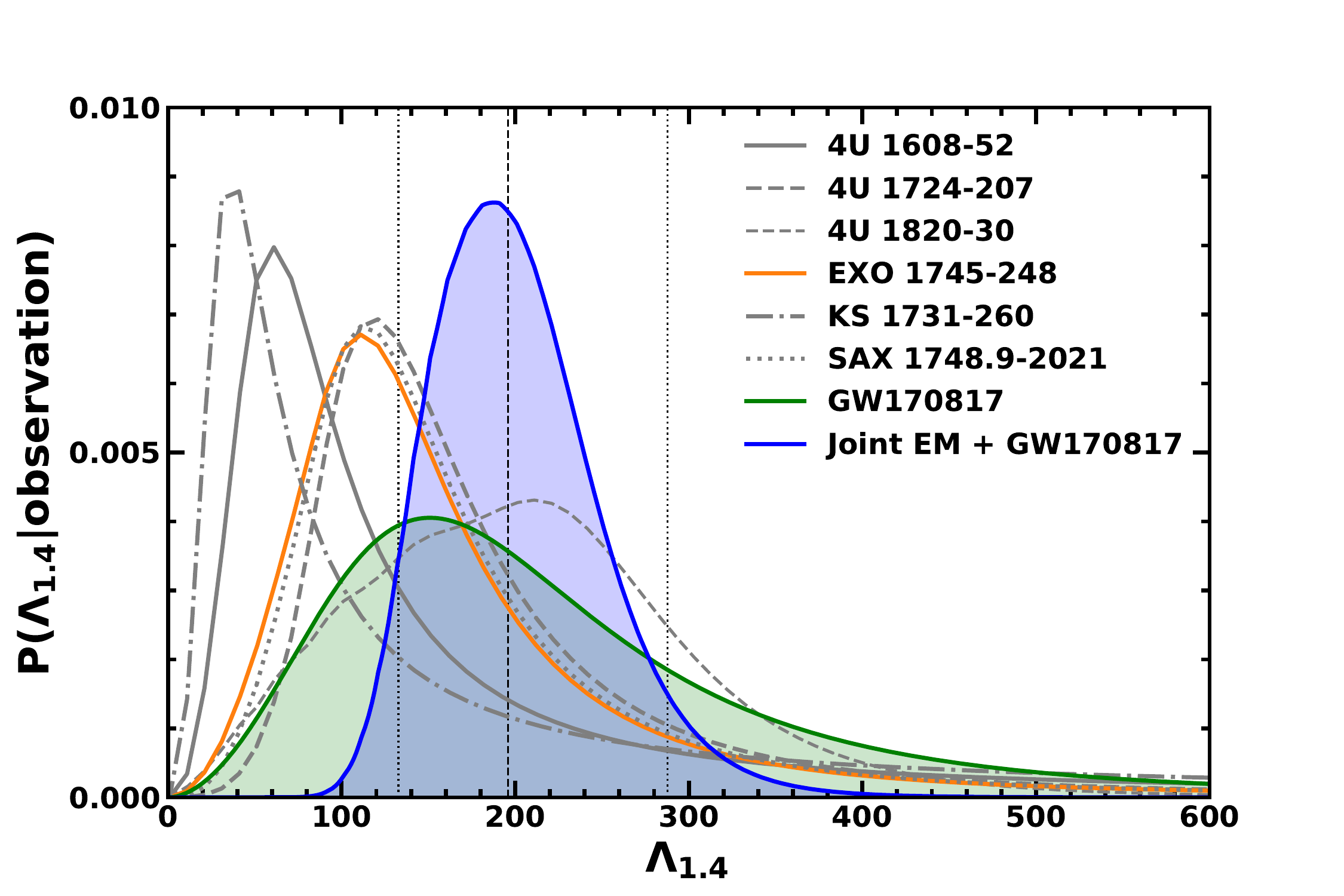}
\caption{Posterior distributions for $\Lambda_{1.4}$. Our model \eqref{model} of the posterior for \citet{LVC_eos}'s $\Lambda_{1.4}$ measurement (green) and the posterior distribution inferred from \citet{Ozel}'s electromagnetic observations of EXO 1745-258 (orange) are shown. The $\Lambda_{1.4}$ posteriors derived from several other observations of thermonuclear bursters are plotted in gray. The combined distribution resulting from the set of electromagnetic observations, plus GW170817, is shown in blue. The median and symmetric 90$\%$ confidence interval of the combined distribution are indicated with the dashed and dotted vertical lines, respectively.}
\label{fig:joint}
\end{figure}

To infer the canonical deformability implied by EXO 1745-248's measured mass and radius, we link $R$ and $\Lambda_{1.4}$ through the universal relations by combining the probability distributions \eqref{biloveprob}, \eqref{iloveprob} and \eqref{icprob}, such that

\begin{equation} \label{Rinf}
P_{\Lambda_{1.4}}(\Lambda_{1.4}|\,\text{EM}) = \frac{G}{c^2} \int \frac{P(M,R|\,\text{EM}) P(GM/c^2 R|\bar{I}) P(\bar{I}|\,\Lambda) P(\Lambda|\,M,\Lambda_{1.4})}{R^2} \, M \, d\Lambda \, d\bar{I} \, dR \, dM .
\end{equation}
This amounts to using the fits \eqref{bilove}, \eqref{ilove} and \eqref{ic} successively to produce a function

\begin{equation} \label{Rmap}
R(M,\Lambda_{1.4}) =\frac{c^2}{GM} \sum_{k=0}^{4} d_k \left[ \sum_{l=0}^{4} c_l \left( \sum_{m=0}^{4} \sum_{n=0}^{1} a_{mn} M^m (\log_{10} \Lambda_{1.4})^n \right)^l \right]^{-k} ,
\end{equation}
while also accounting for the uncertainty in each universal relation.
Equation~\eqref{Rinf} allows us to convert the probability distribution $P(M,R|\,\text{EM})$ constructed from EXO 1745-248's $M$-$R$ posterior samples to a posterior distribution for the canonical deformability, $P_{\Lambda_{1.4}}(\Lambda_{1.4}|\,\text{EM})$. 
This posterior distribution is plotted in Fig.~\ref{fig:joint}.
Calculating its median and symmetric 90$\%$ confidence interval, we find $\Lambda_{1.4} = 139^{+284}_{-82}$.
In other words, the constraint $R_{\text{EM}} = 10.7^{+1.9}_{-1.6}$ stemming from X-ray observations of EXO 1745-248 translates to these bounds on canonical deformability, as the universal relations map the mass-radius posterior $P(M,R|\,\text{EM})$ to the distribution $P_{\Lambda_{1.4}}(\Lambda_{1.4}|\,\text{EM})$ shown in the figure.

We subsequently repeat the EXO 1745-248 analysis for the other {\NS}s listed in Table~\ref{tb:xray}, obtaining posterior distributions $P_{\Lambda_{1.4}}(\Lambda_{1.4}|\,\text{EM}_i)$ for bursters $i=1,...,6$.
We then combine these indirect constraints on $\Lambda_{1.4}$ with the direct measurement from GW170817, $\Lambda_{1.4} = 190^{+390}_{-120}$, to get joint electromagnetic and gravitational-wave constraints that are tighter than the individual measurements.
The combined posterior distribution is computed as

\begin{equation} \label{joint}
P(\Lambda_{1.4}|\,\text{EM,\,GW}) = P(\Lambda_{1.4})\, P(\text{GW}|\,\Lambda_{1.4}) \prod_i P_{\Lambda_{1.4}}(\text{EM}_i|\,\Lambda_{1.4})
\end{equation}
by multiplying the likelihoods $P(\text{GW}|\,\Lambda_{1.4})$ and $P_{\Lambda_{1.4}}(\text{EM}_i|\,\Lambda_{1.4})$ with a chosen prior $P(\Lambda_{1.4})$, lending equal weight to each observation. 
The likelihoods are related to the posteriors by Bayes' theorem:

\begin{equation} \label{lhoods}
P(\text{GW}|\,\Lambda_{1.4}) = \frac{P(\Lambda_{1.4}|\,\text{GW})}{P(\Lambda_{1.4})} , \qquad P_{\Lambda_{1.4}}(\text{EM}_i|\,\Lambda_{1.4}) = \frac{P_{\Lambda_{1.4}}(\Lambda_{1.4}|\,\text{EM}_i)}{P_{\Lambda_{1.4}}(\Lambda_{1.4})}
\end{equation}
up to normalizations. 
The common prior $P_{\Lambda_{1.4}}(\Lambda_{1.4})$ for the electromagnetic observations is calculated from Eq.~\eqref{Rinf} assuming a uniform distribution in $M$ and $R$, i.e.~replacing $P(M,R|\,\text{EM})$ with $P(M,R) = \text{constant}$. 
The mapping \eqref{Rmap} is such that small values of canonical deformability are more likely \textit{a priori}, despite the uninformative mass-radius prior.
The prior $P(\Lambda_{1.4})$ in Eq.~\eqref{joint} is chosen to be identical to the one appearing in Eq.~\eqref{lhoods} for the gravitational-wave observation. 
Then, Eq.~\eqref{joint} reduces to 

\begin{equation}
P(\Lambda_{1.4}|\,\text{EM,\,GW}) = P(\Lambda_{1.4}|\,\text{GW}) \prod_i \frac{P_{\Lambda_{1.4}}(\Lambda_{1.4}|\,\text{EM}_i)}{P_{\Lambda_{1.4}}(\Lambda_{1.4})} ,
\end{equation}
which yields a median and symmetric 90$\%$ confidence interval of $\Lambda_{1.4} = 196^{+92}_{-63}$. 
This joint posterior is plotted in Fig.~\ref{fig:joint}.
As can be seen, the collective impact of the burster measurements is to substantially reduce the size of the error bars on $\Lambda_{1.4}$ relative to the gravitational-wave observation alone; meanwhile, the median is hardly changed. 
This is because most of the electromagnetic mass-radius measurements imply a smaller canonical tidal deformability than GW170817 \textit{a posteriori}, thereby cutting off the long tail of $P(\Lambda_{1.4}|\,\text{GW})$ that extends to large values of $\Lambda_{1.4}$; simultaneously, the bulk of the observations provide minimal support for $\Lambda_{1.4} \lesssim 60$.
Hence, the incorporation of electromagnetic observations of {\NS}s in low-mass X-ray binaries appears to disfavor some of the stiffer candidate \EOSs~that remained compatible with GW170817, while corroborating a canonical deformability of $\approx 200$.

However, we remark that the combined constraint is only as reliable as the simultaneous mass and radius measurements themselves. 
Fig.~\ref{fig:joint} shows that the $\Lambda_{1.4}$ posteriors for 4U 1608-52 and KS 1731-260 are outliers relative to both the GW170817 posterior and the other burster posteriors.
Since $\Lambda_{1.4}$ is a unique property of the \EOS, which is common to all {\NS}s, the discrepancy among maxima \textit{a posteriori} for the electromagnetic measurements indicates that the observations are not, in fact, equally accurate.
As we have not accounted for possible systematic errors in the X-ray observations, it will be interesting to see whether this inference of $\Lambda_{1.4}$ is corroborated by future data from NICER.

\section{Discussion}\label{sec:disc}

In this paper, we used universal relations and constraints on canonical deformability from GW170817 to bound the mass--tidal-deformability, mass--moment-of-inertia and mass-radius relations satisfied by all cold neutron stars. 
We found that the \NS~radius is constrained to be roughly constant for $M \in [1,1.93]\,M_{\odot}$, with radii larger than 13.0 km ruled out at 90$\%$ confidence.
The mass-radius relations that are compatible with GW170817 are also consistent with existing simultaneous mass and radius measurements for six thermonuclear bursters.

Moreover, we inferred tidal deformabilities, moments of inertia, dimensionless spins and radii for individual {\NS}s of interest.
The moments of inertia of a few double {\NS}s were constrained to $\approx 30\%$ accuracy at 90$\%$ confidence by GW170817 and the universal relations, while the canonical \NSx~radius was inferred as $R_{1.4} = 10.9^{+1.9}_{-1.5}$ km.
The dimensionless spins for a set of millisecond pulsars with well-measured masses were calculated to be $\lesssim 0.20$, and those for a set of pulsars in double \NS~systems were found to be $\lesssim 0.04$. 
The spin inferences presented here could be extended to the full population of pulsars with measured masses and rotational frequencies to obtain a spin distribution that is less dependent on \EOSx~modeling. 
The current $\approx 30\%$ level of precision in the inferred spins will improve as more binary \NS~mergers are detected.

The gravitational-wave based predictions for the properties of specific {\NS}s can be compared to direct electromagnetic measurements to test the universality of the {\NS}~\EOS.
Recently, a number of candidate \EOSs~that generically violate the universal relations because of multiple first-order phase transitions or non-standard phases of matter have been proposed \citep{Bandyopadhyay,Han,Lau,Annala17}.
Systematic disagreements between the moments of inertia or radii inferred here and those measured directly via radio or X-ray observations could be interpreted as evidence for such features in the \EOS. 
Alternatively, because the universal relations are different in some modified theories of gravity \citep{Doneva}, a discrepancy could instead indicate support for a modification to general relativity. 

Finally, we investigated how the universal relations can be used to tighten the constraints on $\Lambda_{1.4}$ by combining a gravitational-wave measurement of tidal deformability with electromagnetic observations of {\NS}s in low-mass X-ray binaries. 
Successively employing the binary Love, I-Love and I-compactness relations to create an \EOSx~insensitive $R(M,\Lambda_{1.4})$ relation, we mapped simultaneous mass and radius measurements into posterior probability distributions over $\Lambda_{1.4}$, which were then combined with the corresponding posterior from GW170817. 
Based on the resulting joint distribution, we refined \citet{LVC_eos}'s canonical deformability constraint to $\Lambda_{1.4} = 196^{+92}_{-63}$ at 90$\%$ confidence. 
This inference of $\Lambda_{1.4}$---the most precise to date---is consistent with many (e.g.~\cite{LVC_GW170817,LVC_eos,Annala18}), but not all (e.g.~\cite{Most2018}), previous GW170817-based estimates, and favors a decidedly soft \EOS.

As part of the calculation, we found that the most probable $\Lambda_{1.4}$ values derived from observations of different {\NS}s are not mutually consistent, nor are they all consistent with the canonical deformability implied by GW170817. 
Indeed, the maxima \textit{a posteriori} inferred from observations of 4U 1608-52 and KS 1731-260 are considerably lower than the most probable value indicated by the gravitational-wave event.
Since the derived $R(M,\Lambda_{1.4})$ relation enables us to map disparate radius measurements to a common quantity, $\Lambda_{1.4}$, regardless of the \EOS, and since that quantity can be measured independently using gravitational waves, the joint inference technique presented here may be useful in redressing systematic errors affecting current probes of \NSx~radii.
In any case, additional gravitational-wave observations of binary \NS~mergers and more accurate radius measurements, like those expected from NICER, will permit the universal-relation based bounds on canonical deformability to be further refined.

\acknowledgments

The authors thank Reed Essick and Luciano Rezzolla for helpful discussions about this work, and acknowledge Katerina Chatziioannou for pointing out a mistake in an earlier version of Sec.~\ref{sec:joint}. P.~L.~was supported in part by the Natural Sciences and Engineering Research Council of Canada, and by NSF grants PHY 15-05124 and PHY 17-08081 to the University of Chicago. B.~K.~thanks the Navajbai Ratan Tata Trust, which also provided partial support for this work.

\appendix

\section{Piecewise polytrope parameterizations} \label{sec:pwp}

We calculate piecewise-polytrope fits to the \EOSs~considered here and in \citet{Landry_pulsar} for use in the equations of stellar structure.  
A three-segment piecewise polytrope has been shown to accurately represent a wide range of candidate core \EOSs~\citep{Read}. 
We investigate to what degree the piecewise polytrope parameterization is suitable for unified RMF and SHF \EOSs, and present the best-fit parameter values for the \EOSs~we study.

We adopt the parameterization of \citet{Read}, which approximates the \NSx~\EOS~by a three-segment piecewise polytrope core joined to a low-density crust \EOS. In this model, the \EOS~in the $i^{\text{th}}$ segment is

\begin{equation} \label{pw_p}
p(\rho) = K_i \rho^{\Gamma_i} ,
\end{equation}
where $p$ is the fluid pressure, $\rho$ is the rest-mass energy density, $\Gamma_i$ is the adiabatic index and $K_i$ is a constant of proportionality with units of $(\text{g}/\text{cm}^3)^{1-\Gamma_i}$. 
The dividing densities $\rho_1 = 10^{14.7} \text{g}/\text{cm}^3$, $\rho_2 = 10^{15.0} \text{g}/\text{cm}^3$ between core segments are fixed, so the model has four free parameters: $p_1 = p(\rho_1)$, the pressure at the first dividing density; and $\Gamma_1$, $\Gamma_2$ and $\Gamma_3$, the adiabatic indices for each of the polytropic segments. The model for the crust, based on the SLY4 \EOS, is also fixed. The specification of the four piecewise-polytrope parameters determines the other parameters of the \EOS~recursively---see Appendix A of \citet{Read} for details.

To determine the piecewise-polytrope parameterization for a given unified \EOS, we take its tabulated $p(\rho)$ data and perform a fit to the model described above, minimizing the root-mean-square residual

\begin{equation} \label{res}
\text{res} = \sqrt{\frac{1}{N} \left[ \sum_i \sum_j (\log{p_j} - \log{K_i} + \Gamma_i \log{\rho_j})^2 \right]}
\end{equation}
over the $N$ tabulated data points via a Levenberg-Marquardt algorithm. Here, $i$ labels the piecewise polytrope segments and $j$ labels the data points falling in the density range spanned by the $i^{\text{th}}$ segment. The fit is computed up to the critical density $\rho_{\text{max}}$, the central density that produces the maximum-mass neutron star.

We first repeat the original analysis of \citet{Read} on SLY4, MPA1 and MS1b, which are examples of soft, moderate and stiff \EOSs, respectively. As can be seen by comparing the results in Table~\ref{tb:params1} to Table~III of \citet{Read}, we find comparable values for the fit residual. The fit parameters agree to better than $3\%$. The \NS~properties $M_{\text{max}}$ (maximum mass), $R_{1.4}$ (canonical radius of a $1.4\,M_{\odot}$ star), and $I_{1.338}$ (moment of inertia of a $1.338\,M_{\odot}$ star, like PSR J0737-3039A) are also in good agreement, with $< 1\%$ difference. Having established that our fitting routine is consistent with \citet{Read}'s, we proceed to analyze our unified \EOSs.

The results of the fits are presented in Table~\ref{tb:params1}. The maximum masses computed for the piecewise polytropes are found to agree to within $1\%$ with the values computed for the tabulated \EOSs~in virtually all cases. Similarly, the canonical radii are accurate to better than $1\%$ on average. We remark that the mean error in the maximum mass is smaller for our unified \EOSs~than for those investigated by \citet{Read}; however, the mean error in the canonical radius is larger, while the standard deviation of the error is smaller in both cases. This leads us to conclude that a piecewise polytrope representation of the RMF and SHF \EOSs~is suitable for astrophysical calculations, but that the replacement of the unified crust \EOS~with the fixed SLY4 crust slightly affects the computed radius. Nonetheless, the canonical radius is still recovered to a good approximation. Given the accuracy of the piecewise polytrope models for the unified \EOSs, we adopt this representation for our integrations of the equations of stellar structure.

\setlength{\tabcolsep}{3.8pt}
\LTcapwidth=\textwidth
\begin{center}
\begin{longtable}{l|ccccc|cc|cc|cc}
\caption[Piecewise-polytrope parameterizations for the \EOSs~of interest.]{Piecewise-polytrope parameterizations for the \EOSs~of interest. We report the fit parameters and residual \eqref{res}, as well as several \NS~properties, for each \EOS. The pressure $p_{1}$ is in units of dyne/cm$^2$. The maximum \NS~mass $M_{\text{max}}$ supported by the \EOS, the canonical radius $R_{1.4}$ of a $1.4~M_{\odot}$ \NS, and the double-pulsar moment of inertia $I_{1.338}$ are listed. The \% error in these observables is obtained via $(O_{\text{fit}}/O_{\text{tab}}-1)\times 100$, where $O_{\text{fit}}$ and $O_{\text{tab}}$ are the observables calculated with the best-fit parameterized \EOS~and the tabulated \EOS, respectively. The last two rows give the mean error (ME) and the standard deviation (SD) of the error.} \label{tb:params1} \\
\hline \hline \text{EoS} & $\log _{10}p_1$ & $\Gamma _1$ & $\Gamma _2$ & $\Gamma _3$ & \text{res} & $\left.M_{\max }\text{ [}M_{\odot }\right]$ & $\text{err (\%)}$ & $R_{1.4}\text{ [km]}$ & $\text{err (\%)}$ & $\left.I_{1.338}\text{ [}10^{45}\text{ g }\text{cm}^2\right]$ & $\text{err (\%)}$ \\ \hline
\endfirsthead
\hline \text{EoS} & $\log _{10}p_1$ & $\Gamma _1$ & $\Gamma _2$ & $\Gamma _3$ & \text{res} & $\left.M_{\max }\text{ [}M_{\odot }\right]$ & $\text{err (\%)}$ & $R_{1.4}\text{ [km]}$ & $\text{err (\%)}$ & $\left.I_{1.338}\text{ [}10^{45}\text{ g }\text{cm}^2\right]$ & $\text{err (\%)}$ \\ \hline
\endhead
\hline
\endfoot
\hline \hline
\endlastfoot
\text{BCPM} &34.385 &2.784 &2.920 &2.687 &0.0027 &1.980 &$\phantom{-}$0.016 &11.756 &$-$0.337 &1.280 &$-$0.355\\
\text{BKA20} &34.599 &2.811 &2.451 &1.930 &0.0050 &1.952 &$-$0.196 &13.434 &$\phantom{-}$0.773 &1.590 &$\phantom{-}$0.266 \\
\text{BSk20} &34.377 &3.141 &3.196 &3.042&0.0053 &2.162 &$-$0.195 &11.739 &$\phantom{-}$0.341 &1.301 &$-$0.450\\
\text{BSk21} &34.539 &3.456 &3.073 &2.657 &0.0042 &2.276 &$-$0.065 &12.598 &$\phantom{-}$0.671 &1.475 &$-$0.188\\
\text{BSk22} &34.593 &3.147 &2.865 &2.668 &0.0027 &2.260 &$-$0.172 &13.114 &$\phantom{-}$1.009 &1.558 &$-$0.198\\
\text{BSk23} &34.571 &3.285 &2.954 &2.659 &0.0035 &2.268 &$-$0.106 &12.875 &$\phantom{-}$0.829 &1.520 &$-$0.173\\
\text{BSk24} &34.540 &3.457 &3.072 &2.656 &0.0042 &2.277 &$-$0.061 &12.604 &$\phantom{-}$0.662 &1.476 &$-$0.168\\
\text{BSk25} &34.525 &3.747 &3.067 &2.417 &0.0075 &2.222 &$-$0.055 &12.403 &$\phantom{-}$0.657 &1.449 &$-$0.091\\
\text{BSk26} &34.381 &3.141 &3.193 &3.040 &0.0052 &2.166 &$-$0.177 &11.765 &$\phantom{-}$0.336 &1.305 &$-$0.051\\
\text{BSP} &34.556 &3.204 &2.637 &2.218 &0.0057 &2.022 &$-$0.160 &12.754 &$\phantom{-}$0.667 &1.489 &$\phantom{-}$0.0230\\
\text{BSR2} &34.661 &3.310 &2.951 &2.271 &0.0081 &2.379 &$-$0.148 &13.458 &$\phantom{-}$1.049 &1.638 &$\phantom{-}$0.326\\
\text{BSR2Y} &34.676 &3.378 &2.216 &1.892 &0.0138 &1.993 &$-$0.415 &13.478 &$\phantom{-}$1.521 &1.648 &$\phantom{-}$1.172\\
\text{BSR6} &34.664 &3.028 &3.046 &2.224 &0.0148 &2.422 &$-$0.300 &13.7801 &$\phantom{-}$1.902 &1.681 &$\phantom{-}$0.815\\
\text{BSR6Y} &34.678 &3.075 &2.257 &1.915 &0.0163 &2.018 &$-$0.566 &13.811 &$\phantom{-}$0.893 &1.693 &$\phantom{-}$1.006\\
\text{DD2} &34.638 &3.414 &3.097 &2.322 &0.0141 &2.415 &$-$0.087 &13.234 &$\phantom{-}$0.858 &1.600 &$\phantom{-}$0.302\\
\text{DD2Y} &34.660 &3.523 &2.427 &2.004 &0.0221 &2.087 &$-$0.203 &13.264 &$\phantom{-}$1.203 &1.613 &$\phantom{-}$1.287\\
\text{DDHd} &34.597 &3.573 &2.649 &2.346 &0.0118 &2.125 &$-$0.541 &12.841 &$\phantom{-}$2.197 &1.529 &$-$0.113\\
\text{DDME2} &34.665 &3.639 &3.137 &2.259 &0.0168 &2.482 &$-$0.007 &13.245 &$\phantom{-}$0.589 &1.615 &$\phantom{-}$0.461\\
\text{DDME2Y} &34.679 &3.723 &2.376 &2.081 &0.0194 &2.110&$-$0.135 &13.251&$\phantom{-}$0.752 &1.621 &$\phantom{-}$1.000\\
\text{FSU2} &34.655 &2.675 &2.477 &1.830 &0.0088 &2.068 &$-$0.166 &14.229 &$\phantom{-}$1.135 &1.731 &$\phantom{-}$0.761\\
\text{FSUGarnet} &34.624 &3.538 &2.556 &1.825 &0.0097 &2.063 &$-$0.085 &13.026 &$\phantom{-}$0.829 &1.565 &$\phantom{-}$0.514\\
\text{G3} &34.516 &3.115 &2.735 &2.194 &0.0051 &1.995&$-$0.047 &12.521 &$\phantom{-}$0.091 &1.438 &$-$0.105\\
\text{GM1} &34.679 &2.937 &2.815 &2.438 &0.0031 &2.349 &$-$0.501 &14.019 &$\phantom{-}$2.514 &1.720 &$-$0.397\\
\text{GM1Y} &34.702 &3.032 &2.716 &2.013 &0.0126 &1.980 &$-$0.608 &14.063 &$\phantom{-}$2.862 &1.740 &$\phantom{-}$0.741\\
\text{IOPB} &34.640 &3.253 &2.664 &1.786 &0.0141 &2.147 &$-$0.038 &13.354 &$\phantom{-}$0.368 &1.614 &$\phantom{-}$0.997\\
\text{KDE0v1} &34.366 &2.791 &2.897 &2.779 &0.0049 &1.967 &$-$0.081 &11.586 &$-$0.072 &1.250 &$-$0.310\\
\text{Model1} &34.601 &3.247 &2.560 &1.830 &0.0094 &2.012 &$-$0.022 &13.053 &$\phantom{-}$0.425 &1.552 &$\phantom{-}$0.506\\
\text{MPA1} &34.477 &3.441 &3.580 &2.884 &0.0078 &2.434 &$-$0.912&12.343 &$-$1.250 &1.429 &$-$0.444\\
\text{MS1b} &34.845 &3.410 &3.030 &1.467 &0.0154 &2.736 &$-$1.647 &14.535 &$-$0.645 &1.870 &$-$2.276\\
\text{NL3} &34.847 &3.246 &3.098 &1.298 &0.0237 &2.759 &$-$0.540 &14.810 &$\phantom{-}$1.799 &1.916 &$\phantom{-}$0.938\\
\text{NL3$\omega\rho$} &34.821 &3.974 &3.127 &1.552 &0.0202 &2.745 &$-$0.240 &13.796 &$\phantom{-}$0.745 &1.744 &$\phantom{-}$0.621\\
\text{NL3$\omega\rho$Y} &34.809 &3.922 &2.264 &2.166 &0.0120 &2.334 &$-$0.292 &13.773 &$\phantom{-}$0.579 &1.713 &$-$1.117\\
\text{NL3$\omega\rho$Yss} &34.805 &3.913 &1.895 &2.106 &0.0141 &2.138 &$-$0.260 &13.735 &$\phantom{-}$0.502 &1.642 &$-$5.308\\
\text{NL3Y} &34.810 &3.092 &2.222 &2.214 &0.0092 &2.303 &$-$1.049 &14.813 &$\phantom{-}$1.768 &1.903 &$\phantom{-}$0.216\\
\text{NL3Yss} &34.802 &3.062 &1.766 &2.051 &0.0118 &2.058 &$-$0.496 &14.812 &$\phantom{-}$1.767 &1.900 &$\phantom{-}$0.055\\
\text{Rs} &34.555 &2.674 &2.670 &2.670 &0.0017 &2.104&$-$0.584&13.219 &$\phantom{-}$2.568 &1.532 &$-$0.845\\
\text{SINPA} &34.593 &3.321 &2.563 &1.839 &0.0088 &1.999 &$-$0.064 &12.941 &$\phantom{-}$0.544 &1.535&$\phantom{-}$0.408\\
\text{SK255} &34.549 &2.623 &2.758 &2.703 &0.0031 &2.138 &$-$0.253 &13.245 &$\phantom{-}$1.099 &1.531 &$-$0.625\\
\text{SK272} &34.574 &2.730 &2.848 &2.766 &0.0037 &2.225 &$-$0.245&13.370 &$\phantom{-}$0.766 &1.568&$-$0.475\\
\text{SKa} &34.546 &2.810 &2.873 &2.783 &0.0026 &2.202 &$-$0.276 &13.031 &$\phantom{-}$1.209 &1.512 &$-$0.594\\
\text{SKb} &34.507 &3.143 &2.909 &2.808 &0.0047 &2.174 &$-$0.630 &12.497 &$\phantom{-}$2.675 &1.437 &$-$0.744\\
\text{SkI2} &34.613 &2.658 &2.588 &2.649 &0.0033 &2.149 &$-$0.614 &13.825 &$\phantom{-}$2.893 &1.648 &$-$0.796\\
\text{SkI3} &34.632 &2.824 &2.676 &2.697 &0.0027 &2.230 &$-$0.397 &13.765 &$\phantom{-}$1.911 &1.657 &$-$0.473\\
\text{SkI4} &34.507 &3.111 &2.909 &2.734 &0.0024 &2.161 &$-$0.340 &12.517 &$\phantom{-}$1.460 &1.439 &$-$0.480\\
\text{SkI5} &34.663 &2.587 &2.572 &2.718 &0.0043 &2.224 &$-$0.690 &14.520 &$\phantom{-}$3.502 &1.776 &$-$0.862\\
\text{SkI6} &34.519 &3.107 &2.918 &2.734 &0.0020 &2.183 &$-$0.287 &12.611 &$\phantom{-}$1.272 &1.457 &$-$0.415\\
\text{SkMP} &34.508 &2.782 &2.777 &2.729 &0.0007 &2.096 &$-$0.489 &12.699 &$\phantom{-}$1.941 &1.447 &$-$0.756\\
\text{SKOp} &34.451 &2.672 &2.712 &2.635 &0.0015 &1.966 &$-$0.321 &12.228 &$\phantom{-}$1.143 &1.350 &$-$0.645\\
\text{SLY230a} &34.399 &3.150 &3.082 &2.789 &0.0038 &2.093 &$-$0.237 &11.821 &$\phantom{-}$0.174 &1.314 &$-$0.357\\
\text{SLY2} &34.392 &2.959 &2.984 &2.829 &0.0041 &2.042 &$-$0.538 &11.777 &$\phantom{-}$0.220 &1.295 &$-$0.426\\
\text{SLY4} &34.380 &2.979 &2.999 &2.849 &0.0040 &2.048 &$-$0.092 &11.700 &$\phantom{-}$0.231 &1.282 &$-$0.410\\
\text{SLY9} &34.493 &2.992 &2.936 &2.750 &0.0027 &2.153 &$-$0.109 &12.485 &$\phantom{-}$0.441 &1.425 &$-$0.380\\
\text{TM1} &34.701 &2.754 &2.472 &1.870 &0.0067 &2.169 &$-$0.479 &14.540 &$\phantom{-}$2.282 &1.806 &$\phantom{-}$0.025\\
\text{Mean} & &  &  &  &  &  &$-$0.368 &  &$\phantom{-}$1.324&  &$-$0.139\\
\text{Std Dev} &  &  &  &  &  &  &$\phantom{-}$0.480 &  &$\phantom{-}$1.278 &  &$\phantom{-}$0.949\\
\end{longtable}
\end{center}

\vspace{-1.32cm}

\bibliography{universalnsprops-refs}

\end{document}